\documentclass[12pt]{article}

\usepackage{amsmath}
\usepackage{subfigure}
\usepackage{appendix,epic,eepic,amsmath,amsthm,amssymb,epsfig,cite,indentfirst,oldgerm}
\renewcommand{\theequation}{\arabic{equation}}
\newcommand{\EQ}{\begin{equation}}
\newcommand{\EN}{\end{equation}}

\newcommand{\bear}{\begin{eqnarray}}
\newcommand{\ear}{\end{eqnarray}}
\newcommand{\bt} { \begin{tabular} }
\newcommand{\et}{ \end{tabular} }
\newcommand{\bc} { \begin{center} }
\newcommand{\ec}{ \end{center} }

\newcommand{\btb} { \begin{table} }
\newcommand{\etb}{ \end{table} }

\begin{document}

\topmargin 0pt
\oddsidemargin 5mm
\newcommand{\NP}[1]{Nucl.\ Phys.\ {\bf #1}}
\newcommand{\PL}[1]{Phys.\ Lett.\ {\bf #1}}
\newcommand{\NC}[1]{Nuovo Cimento {\bf #1}}
\newcommand{\CMP}[1]{Comm.\ Math.\ Phys.\ {\bf #1}}
\newcommand{\PR}[1]{Phys.\ Rev.\ {\bf #1}}
\newcommand{\PRL}[1]{Phys.\ Rev.\ Lett.\ {\bf #1}}
\newcommand{\MPL}[1]{Mod.\ Phys.\ Lett.\ {\bf #1}}
\newcommand{\JETP}[1]{Sov.\ Phys.\ JETP {\bf #1}}
\newcommand{\TMP}[1]{Teor.\ Mat.\ Fiz.\ {\bf #1}}

\renewcommand{\thefootnote}{\fnsymbol{footnote}}

\newpage
\setcounter{page}{0}
\begin{titlepage}
\begin{flushright}

\end{flushright}
\vspace{0.5cm}
\begin{center}
{\large An Integrable Nineteen Vertex Model Lying on a Hypersurface}\\
\vspace{1cm}
{\large M.J. Martins } \\
\vspace{0.15cm}
{\em Universidade Federal de S\~ao Carlos\\
Departamento de F\'{\i}sica \\
C.P. 676, 13565-905, S\~ao Carlos (SP), Brazil}\\
\vspace{0.35cm}
\end{center}
\vspace{0.5cm}

\begin{abstract}
We have found 
a family of solvable nineteen vertex model with statistical
configurations invariant by the time reversal symmetry
within a systematic study of the respective Yang-Baxter relation.
The Boltzmann weights sit on a degree seven
algebraic threefold which is shown birationally equivalent 
to the three-dimensional projective space. This permits to write
parameterized expressions for both the transition operator
and the $\mathrm{R}$-matrix depending on three independent
affine spectral parameters. The Hamiltonian limit
tells us that the azimuthal magnetic field term
is connected with the asymmetry among 
two types of spectral variables. The absence
of magnetic field defines a physical submanifold
whose geometrical properties are remarkably shown
to be governed by a quartic $\mathrm{K}3$ surface. 
This expands considerably the class of irrational manifolds
that could emerge in the theory of quantum 
integrable models.
\end{abstract}

\vspace{.15cm} \centerline{}
\vspace{.1cm} \centerline{Keywords: Yang-Baxter Equation, Vertex Models, Algebraic Geometry}
\vspace{.15cm} \centerline{October,~2014}

\end{titlepage}


\pagestyle{empty}

\newpage

\pagestyle{plain}
\pagenumbering{arabic}

\renewcommand{\thefootnote}{\arabic{footnote}}
\newtheorem{proposition}{Proposition}
\newtheorem{pr}{Proposition}
\newtheorem{remark}{Remark}
\newtheorem{re}{Remark}
\newtheorem{theorem}{Theorem}
\newtheorem{theo}{Theorem}

\def\ll{\left\lgroup}
\def\rr{\right\rgroup}

\newtheorem{Theorem}{Theorem}[section]
\newtheorem{Corollary}[Theorem]{Corollary}
\newtheorem{Proposition}[Theorem]{Proposition}
\newtheorem{Conjecture}[Theorem]{Conjecture}
\newtheorem{Lemma}[Theorem]{Lemma}
\newtheorem{Example}[Theorem]{Example}
\newtheorem{Note}[Theorem]{Note}
\newtheorem{Definition}[Theorem]{Definition}

\section{Introduction}

At present the 
method of commuting transfer matrices provides the most important
device for constructing exactly
solvable lattice systems of statistical 
mechanics in two spatial dimensions\cite{BAX}. 
Let us denote by 
$\mathrm{T}_{\mathrm{N}}(\omega)$ the model transfer matrix defined on
a given direction of the lattice with 
length $\mathrm{N}$.
In order to make notation simpler we have represented
the lattice Boltzmann weights 
$\omega_1,\cdots,\omega_m$ by the single vector
$\omega \in \mathbb{C}^m$. 
The commutativity of two transfer matrices with distinct weights
implies the condition, 
\EQ
\left[ \mathrm{T}_{\mathrm{N}}(\omega^{'}),
\mathrm{T}_{\mathrm{N}}(\omega^{''}) \right]=0,
~~\forall~\omega^{'}~\mathrm{and}~\omega^{''},
\label{comm}
\EN
for arbitrary length $\mathrm{N}$.

The fact that such commutation relation 
depends on the size $\mathrm{N}$ seems that one needs to verify 
an infinite number of relations
among the Boltzmann weights to conclude that two different transfer matrices 
indeed commute.
This is fortunately not the case  since
Baxter \cite{BAX} argued that 
it is sufficient to solve only a finite set of algebraic relations 
to built up a family of commuting transfer matrices 
for any size $\mathrm{N}$. 
This local condition is often
referred as Yang-Baxter equation and its specific structure depends much on the
class of lattice system under consideration. In this paper we are interested to
investigate novel solutions to this relation in the case of lattice vertex models.
We
recall that the fluctuation 
variables of vertex models lie on the bonds 
between neighboring lattice points and the interaction
energies depend on the allowed vertices configurations.
The main feature of these models are their inherent 
tensor structure which allows us to construct the corresponding transfer matrices
out of a single local transition operator. 
In the simplest case of rectangular lattices this operator acts on the
direct product of the auxiliary and quantum spaces associated respectively
to the horizontal and vertical edges statistical configurations.
Assuming that 
each edge of the rectangular lattice can 
take values on $\mathrm{q}$ possible states one can represent the
transition operator $\mathrm{L}(\omega)$ on the auxiliary space as the following
$\mathrm{q} \times \mathrm{q}$ matrix, 
\EQ
{\mathrm{L}}(\omega)=\left(\begin{array}{c|c|c|c}
                {\mathrm W}_{1,1} & {\mathrm W}_{1,2} & \cdots & {\mathrm W}_{1,q} \\ \hline
                {\mathrm W}_{2,1} & {\mathrm W}_{2,2} & \cdots & {\mathrm W}_{2,q} \\ \hline
                \vdots & \vdots & \ddots & \vdots \\ \hline
                {\mathrm W}_{q,1} & {\mathrm W}_{q,2} & \cdots & {\mathrm W}_{q,q} \\ 
                \end{array}\right).
\label{mono}
\EN

The entries
${\mathrm  W}_{a,b}$ are also
$\mathrm{q} \times \mathrm{q}$ operators but now acting
on the space
of quantum vertical degrees of freedom. Their matrix elements 
${\mathrm  W}_{a,b}(c,d)$ represent the Boltzmann weights 
for the edge horizontal states
$a,b$ and the edge vertical configurations $c,d$.
The maximum number $m$ of
distinct Boltzmann weights the vertex model can have is therefore
$m=\mathrm{q}^4$.

The transition operators can be combined to construct for instance the 
row-to-row transfer matrix represented as operators in the quantum
space variables with an arbitrary number $\mathrm{N}$ of columns. Considering
periodic boundary conditions on the horizontal direction the transfer matrix
takes the form,
\EQ
\mathrm{T}_{\mathrm{N}}(\omega)=\mathrm{Tr}_{\mathrm{q}} 
\left[ \mathrm{L}_1(\omega) \mathrm{L}_2(\omega) \cdots \mathrm{L}_{\mathrm{N}}(\omega) \right ],
\label{trans}
\EN
where the matrix multiplication and the 
trace operations are performed on the
auxiliary space. The subscript index for 
the transition operator  
$\mathrm{L}_j(\omega)$ means that its
matrix elements act non-trivially only at the
$\mathrm{j}$-$\mathrm{th}$ vertical quantum space of states. 

A sufficient condition 
for the commutativity 
of $\mathrm{T}_{\mathrm{N}}(\omega^{'})$ 
and $\mathrm{T}_{\mathrm{N}}(\omega^{''})$ 
assumes the existence
of a non-singular
$\mathrm{q}^2 \times \mathrm{q}^2$ numerical matrix
$\mathrm{R}(\mathbf{w})$ which together with the 
transition operator fulfill the renowned Yang-Baxter
relation \cite{BAX}, 
\EQ
\mathrm{R}(\mathbf{w}) \left[\mathrm{L}(\omega^{'}) \otimes \mathrm{I}_{\mathrm{q}}\right]
\left[\mathrm{I}_{\mathrm{q}} \otimes \mathrm{L}(\omega^{''})\right] =
\left[\mathrm{I}_{\mathrm{q}} \otimes \mathrm{L}(\omega^{''})\right] \left[\mathrm{L}(\omega^{'}) 
\otimes \mathrm{I}_{\mathrm{q}} \right]
\mathrm{R}(\mathbf{w}),
\label{YBA}
\EN
where $\mathrm{I}_{\mathrm{q}}$ denotes the  
$\mathrm{q} \times \mathrm{q}$ unity matrix and 
the tensor product is considered within the auxiliary
space.
We have used the 
bold symbol $\mathbf{w}$ to emphasize that the entries 
of the $\mathrm{R}$-matrix should not be
confused with the set of Boltzmann 
weights $\omega$ defining the transition operator. 

Nowadays we are aware of several series of integrable 
vertex models for every value
of $\mathrm{q}$ and different types of statistical
configurations. 
Notable examples are the lattice models 
with trigonometric weights associated to the 
representation theory of deformed Lie algebras \cite{JI,BAZ}
and their
elliptic generalizations \cite{BAX1,BELA}. 
However, 
we still do not know any criterion to find the most general 
statistical  
configurations of the Boltzmann weights for which
the existence of non-trivial solutions of the 
Yang-Baxter equation could be assured. Even for a given lattice 
statistical configuration it is an open 
problem to describe by what
means the explicit form of all possible corresponding matrices
${\mathrm{L}}(\omega)$ and 
$\mathrm{R}(\mathbf{w})$ 
would be obtained. This latter question is equivalent
to find the irreducible zeroes sets of a large number of 
homogeneous polynomials with many distinct monomials arising
from the Yang-Baxter equation (\ref{YBA}). This is the typical problem 
one faces in Algebraic Geometry which in our case is formulated on 
the product of three projective spaces denoted here by
$\mathbb{CP}^{m-1} \times 
\mathbb{CP}^{m-1} \times \mathbb{CP}^{m-1}$. 
To the best of our knowledge it
is not clear how the wonderful results in this 
field of mathematics could
be used to shed some light into 
the theory of classification
of solutions 
of the Yang-Baxter equation. In fact, it appears that so far some
basic statements using Algebraic Geometry methods 
has been restricted to   
vertex models with two states per edge yet under additional
assumption on the transition operator properties \cite{KRI}.
In spite of that we can still sketch
some practical guidelines for searching solutions of the Yang-Baxter
relation in the realm of Algebraic Geometry. We hope
that this approach  could be useful 
at least for vertex models with
a specific given statistical configuration and for moderate number
of edge states. 
These basic points to be described below are certainly influenced by
the celebrated analysis of the Yang-Baxter relation  for
the eight-vertex model \cite{BAX1}. The manner we elaborate upon this method has
been however inspired in our previous experience in 
dealing with several types of functional relations associated 
to a three-state vertex model \cite{MAR}.

We start by recalling that the number of 
functional relations coming from Eq.(\ref{YBA})
is generically much larger than
the corresponding number of Boltzmann weights we need to determine.
This leads us to solve
a very overdetermined system of homogenous polynomial equations which
are however linear on the entries of the $\mathrm{R}$-matrix. We 
can use this feature to fix the basic structure of the matrix
$\mathrm{R}(\mathbf{w})$ by means of standard linear elimination of its
elements out of a suitable subset of independent functional relations.
The $\mathrm{R}$-matrix will be ultimately dependent on the 
set of weights 
$\omega^{'}$ and $\omega^{''}$ and this means that 
we can formally rewrite the Yang-Baxter
equation as, 
\EQ
\mathrm{R}(\omega^{'},\omega^{''}) \left[ \mathrm{L}(\omega^{'}) \otimes \mathrm{I}_{\mathrm{q}} \right]
\left[\mathrm{I}_{\mathrm{q}} \otimes \mathrm{L}(\omega^{''})\right] =
\left[\mathrm{I}_{\mathrm{q}} \otimes \mathrm{L}(\omega^{''})\right] \left[\mathrm{L}(\omega^{'}) 
\otimes \mathrm{I}_{\mathrm{q}}\right]
\mathrm{R}(\omega^{'},\omega^{''}).
\label{YBN}
\EN

The fact that 
$\mathrm{T}_{\mathrm{N}}(\omega)$ always commutes with itself
should be encoded as particular solution of the Yang-Baxter equation for
general transition operators. Direct inspection of
Eq.(\ref{YBN}) at the point 
$\omega^{'}=\omega^{''}$ tells us that the $\mathrm{R}$-matrix simply switches
the order of the tensor product of two transition operators with equal weights.
We then conclude that such trivial solution for arbitrary 
transition operators is attained
imposing the initial condition,
\EQ
\mathrm{R}(\omega,\omega)=\xi(\omega)\mathrm{P}_{\mathrm{q}}, 
\label{INI}
\EN
where $\xi(\omega)$ is a normalization and the operator
$\mathrm{P}_{\mathrm{q}}$  denotes
the $\mathrm{q}^{2} \times \mathrm{q}^{2}$ permutator.  

We next note that the Yang-Baxter relation (\ref{YBN}) 
provides us certain consistent
condition on the $\mathrm{R}$-matrix upon exchange of the weights 
$\omega^{'}$ and $\omega^{''}$. 
In order to see that we first interchange single primed and 
double primed weights labels in Eq.(\ref{YBN})
and afterwards we use the help of the permutator 
$\mathrm{P}_{\mathrm{q}}$ 
to reorder  
the tensor product of two different transition operators. As a final result we obtain,
\EQ
\mathrm{P}_{\mathrm{q}}\mathrm{R}(\omega^{''},\omega^{'})\mathrm{P}_{\mathrm{q}} 
\left[\mathrm{I}_{\mathrm{q}} \otimes \mathrm{L}(\omega^{''}) \right]
\left[\mathrm{L}(\omega^{'}) \otimes \mathrm{I}_{\mathrm{q}}\right] =
\left[\mathrm{L}(\omega^{'}) \otimes \mathrm{I}_{\mathrm{q}}\right] 
\left[\mathrm{I}_{\mathrm{q}} \otimes \mathrm{L}(\omega^{''}) \right]
\mathrm{P}_{\mathrm{q}}\mathrm{R}(\omega^{''},\omega^{'})\mathrm{P}_{\mathrm{q}},
\label{YBN1}
\EN

The above expression can be further simplified once we apply 
on its left hand side the $\mathrm{R}$-matrix 
$\mathrm{R}(\omega^{'},\omega^{''})$  and after that we use
the Yang-Baxter equation (\ref{YBN}) 
to rearrange the order of the right hand side transition operators. Considering
these manipulations we find that Eq.(\ref{YBN1}) can be rewritten as
the following commutator,
\EQ
\left[ \mathrm{R}(\omega^{'},\omega^{''}) 
\mathrm{P}_{\mathrm{q}}\mathrm{R}(\omega^{''},\omega^{'})\mathrm{P}_{\mathrm{q}},
[\mathrm{I}_{\mathrm{q}} \otimes \mathrm{L}(\omega^{''})]
[\mathrm{L}(\omega^{'}) \otimes \mathrm{I}_{\mathrm{q}}] \right] =0.
\label{comm1}
\EN

The left term of the commutator (\ref{comm1}) is a scalar on the tensor product of two
auxiliary spaces  whereas 
the right term for an arbitrary 
$\mathrm{L}(\omega)$ turns out to be a complicated operator 
in this same space. This means that a generic solution to Eq.(\ref{comm1})
occurs when its left term becomes proportional to the identity matrix 
in the product of auxiliary spaces. This leads us to what is usually called 
unitarity condition or inversion relation \cite{STRO}
for the $\mathrm{R}$-matrix, namely 
\EQ
\mathrm{R}(\omega^{'},\omega^{''}) 
\mathrm{P}_{\mathrm{q}} \mathrm{R}(\omega^{''},\omega^{'})
\mathrm{P}_{\mathrm{q}}=  
\rho(\omega^{'},\omega^{''})  \mathrm{I}_{\mathrm{q}} \otimes \mathrm{I}_{\mathrm{q}},
\label{UNI}
\EN
where $\rho(\omega^{'},\omega^{''})$ is a scalar normalization. We observe that 
unitarity of the $\mathrm{R}$-matrix is fully compatible with the initial condition (\ref{INI}).

From now on we shall assume that we are 
dealing with integrable
vertex models whose $\mathrm{R}$-matrices 
satisfy the unitarity property (\ref{UNI}) which assures 
us  that they are invertible. At this point 
it is natural to ask whether or not 
the unitarity relation together with the Yang-Baxter equation
are capable to impose any relevant restriction on
the functional equations when their single and
double primed weights
labels are exchanged. This idea has already been explored
in our recent work \cite{MAR} and  
we shall here present only the
main conclusion.
Let $\mathrm{F}_j(\omega^{'},\omega^{''})$  be
the polynomials derived from
the Yang-Baxter equation after we have performed
the elimination of
the $\mathrm{R}$-matrix
elements. It has been shown that such polynomials 
have to satisfy the following 
anti-symmetrical property, 
\EQ
\mathrm{F}_j(\omega^{'},\omega^{''})+
\mathrm{F}_j(\omega^{''},\omega^{'})=0.
\label{ANTI}
\EN

For a general vertex model it is not expected 
that all the functional equations   
to be satisfied are
automatically anti-symmetrical upon the exchange of
the weights
$\omega^{'}$ and $\omega^{''}$.  In fact, the
requirement that  
$\mathrm{F}_j(\omega^{'},\omega^{''})$ should satisfy 
the property (\ref{ANTI}) has been decisive to simplify 
cumbersome high degree polynomials expressions emerging
in the analysis of a three-state vertex model \cite{MAR}.

In order to make further
progress it is crucial that we are able to recast 
at least part of the anti-symmetrical polynomials 
$\mathrm{F}_j(\omega^{'},\omega^{''})$
in the following 
particular factorized form,
\EQ
\mathrm{F}_j(\omega^{'},\omega^{''})=
\mathrm{H}_j(\omega^{'})
\mathrm{G}_j(\omega^{''})-
\mathrm{H}_j(\omega^{''})
\mathrm{G}_j(\omega^{'}),~~j=1,\cdots,\mathrm{n},
\label{fact}
\EN
for some integer $\mathrm{n}$. The homogeneous polynomials 
$\mathrm{H}_j(\omega)$ and 
$\mathrm{G}_j(\omega)$ are assumed to be irreducible
having the same degree  
on the weights.

The above step provides us the basic ingredient to start
the construction of two commuting transfer matrices whose
weights will be sited on the same algebraic variety.
This can be achieved imposing that
each factorized functional relation (\ref{fact})
vanishes upon
the choice of the same polynomial
restriction for both set of
variables $\omega^{'}$ and $\omega^{''}$. Such special
solution to Eq.(\ref{fact}) in which the weights with
distinct labels are separated is clearly given by,
\EQ
\frac{\mathrm{H}_j(\omega)}{
\mathrm{G}_j(\omega)}=\Lambda_j 
,~~j=1,\cdots,\mathrm{n},
\label{solD}
\EN
where $\Lambda_1,\cdots,\Lambda_{\mathrm{n}}$ are free parameters. 

In the language of
Algebraic Geometry the
particular solution (\ref{solD}) can be seen as a prime divisor over the
algebraic set made out of the zeroes of the factorized 
polynomial we have started with. For the technical details 
concerning this interpretation
see \cite{MAR} and in what follows we shall 
refer to such special solutions as divisors. 

We next have to deal with the remaining functional relations
which could not be brought into 
the suitable factorized form (\ref{fact}). In these cases
we hope that such polynomials can also be set equal to zero 
at the expense of
imposing additional constraints
on the available free parameters $\Lambda_j$. 
Being successful on such last step we then have solutions
of the Yang-Baxter equation whose properties are formally governed
by the algebraic variety,
\EQ
\label{vari}
\mathrm{Y}=\{\omega
\in \mathbb{CP}^{m-1} | \mathrm{H}_1(\omega)-\Lambda_1 \mathrm{G}_1(\omega)=0, 
\mathrm{H}_2(\omega)-\Lambda_2 \mathrm{G}_2(\omega)=0,\cdots, 
\mathrm{H}_{\mathrm{n}}(\omega)-\Lambda_{\mathrm{n}} \mathrm{G}_{\mathrm{n}}(\omega)=0 
\}, 
\EN
where now a subset of the parameters 
$\Lambda_1,\cdots,\Lambda_{\mathrm{n}}$ may be fixed.

The complete characterization of the geometrical features
of the projective variety $\mathrm{Y}$ certainly depends 
much on the polynomial form of 
the generators of its ideal. For example, the intersection of many polynomials can in principle
give rise to a number of irreducible varieties none of which being superfluous.
However, there exists an important invariant of the variety $\mathrm{Y}$
for which we can make a concrete statement without the knowledge of the specific structure
of the polynomials (\ref{solD}).  
This turns out to be the maximal of the dimensions of
the irreducible components of $\mathrm{Y}$ denoted here by the symbol $\mathrm{dim}(\mathrm{Y})$.
The Algebraic Geometry theory predicts a lower bound for such invariant which is \cite{HAR},
\EQ
\mathrm{dim}(\mathrm{Y}) \geq (\mathrm{m}-1)-\mathrm{n},
\label{DIM}
\EN
and when the equality holds such component of $\mathrm{Y}$ is named 
a complete intersection. 

The dimension of the variety underlying a solution of the Yang-Baxter
equation dictates the number of free weights or spectral parameters 
expected to be present in
the uniformization of the respective transition operator.   
We believe that it is of great interest to search for 
integrable systems whose weights
lie on high dimensional algebraic varieties. For instance
even a rational three-dimensional variety can contain several non-rational 
surfaces and some of them could still represent 
a submanifold of physical
interest. This study provides us a clear route to discover 
examples of solvable models lying 
on irrational varieties more involved than those uniformized by high genus curves  
such as the chiral Potts model \cite{CHI1,CHI2}. In fact, we are not aware of
examples of integrable models with weights lying on non-rational surfaces 
which are not ruled by algebraic curves.

The main purpose of this paper
is to investigate the above possibility in the case of a rather general three-state
vertex model with ice-rule statistical configurations. 
The corresponding transition
operator commutes with the azimuthal component of the spin-$1$ generators
and this degree of freedom could somewhat be interpreted as the presence
of an extra spectral parameter.
In general, we expect that generic solvable $\mathrm{U}(1)$ invariant 
vertex models will contain the minimum number of 
two free spectral variables and consequently 
their weights should at least be sited on two-dimensional
algebraic manifolds. Another motivation comes from the existence 
of a solvable spin-$1$ quantum chains with three main free coupling constants
discovered by Crampr\'e, Frappat and Ragoucy \cite{CFR}
within the coordinate Bethe ansatz. The respective spin-$1$ Hamiltonian is
a generalization of the one built out of colored 
transition operators based on representations 
of the algebra $\mathrm{U[SU(2)]_q}$ 
when $\mathrm{q}$ is at roots of unity \cite{DEG,GOM}.
It is therefore conceivable that some of the Hamiltonian couplings
could originate through the presence of additional spectral variables rather 
than from the usual constants associated to divisors. 
We found that this is indeed the situation of the spin chain denominated
$\mathrm{SpR}$ in the reference \cite{CFR}. We shall
show that the  integrability properties are governed by
a transition operator sitting on a degree seven 
algebraic threefold with polynomial
coefficients depending on two arbitrary constants. As a result we then  
have three independent spectral parameters at our disposal 
and one of them reflects the solvability of the Hamiltonian
in the presence of any azimuthal  magnetic field. 
Remarkably enough, the submanifold giving rise to
the spin-$1$ Hamiltonian in absence of the magnetic field is governed by
the geometric properties of an algebraic surface on the $\mathrm{K}3$ class.
Recall here that $\mathrm{K}3$ surfaces 
have zero Kodaira dimension\footnote{There exists a rough relationship 
between positive(negative) Kodaira dimension 
and the negative(positive) curvature of the surface. A zero value
for the Kodaira dimension corresponds to flatness 
and for details of definitions and properties 
we refer to the book \cite{BEAU}.}
being two-dimensional
Calabi-Yau manifolds which do not have a group structure.
This appears to be the
first example of a solution of the 
Yang-Baxter equation lying on such 
famous family of compact complex surfaces.

We have organized this paper as follows. In next Section we present the 
structure of the transition
operator of the nineteen vertex model. We assume that the 
model is invariant by time reversal symmetry
leading us to the parameter subspace of fourteen non-null weights. The
analysis of the functional relations is performed in Section \ref{sectionYBE} 
and we find an integrable vertex model sitting on an algebraic threefold with
two free couplings. In Section \ref{sectionGEOM} we show that such threefold
is birationally equivalent to the projective space $\mathbb{CP}^{3}$. This
mapping is used to present the parameterized form of the respective transition
operator. We discuss the Hamiltonian limit of the vertex model and some
of its couplings originate from combinations 
of three independent spectral weights.
In particular the presence
of an arbitrary magnetic field is related to the asymmetry 
of two types of weights. In Section \ref{sectionSUBMANI} we discuss the
the submanifold associated with the absence of
any magnetic field and we show 
that its geometrical properties are governed by $\mathrm{K}3$ surfaces. 
The expression of the respective transition
operator lying on a quartic $\mathrm{K}3$ surface with only canonical
singularities is provided. We have summarized our concluding remarks in
Section \ref{sectionCON}. In three Appendices we have presented 
certain technical details omitted in the main text,
the expressions of the
$\mathrm{R}$-matrices on three different embeddings 
as well as the computations of the Hamiltonian limit.

\section{The Nineteen Vertex Model}
\label{sectionVERTEX}

The nineteen vertex model has
three states per bond and its statistical
configurations are restricted by the ice rule. This means that
the weights $\mathrm{W}_{a,b}(c,d)$ are non null only
when the state variables at the vertex 
satisfy the condition $a+c=b+d$. Here we
shall consider a subclass of such models 
whose Boltzmann weights are invariant when we rotate the lattice
of 180 degrees. In analogy with relativistic 1+1 dimensional scattering
theory \cite{ZAMO} this invariance is often  denominated
time reversal symmetry, 
\EQ
\mathrm{W}_{a,b}(c,d)=\mathrm{W}_{b,a}(d,c).
\label{TIME}
\EN

We would like to remark that the request of time reversal invariance forces us 
from the very beginning to be far away of the 
recent found integrable genus five manifold \cite{MAR}.
We also note that this symmetry is not that stringent since the vertex model space of parameters
is reduced to still fourteen distinct weights. These facts favor the possibility
of uncovering new solvable nineteen vertex models which hopefully will be sited on high
dimensional varieties. The explicit 
distinct matrix elements of the 
transition operator are,
\begin{eqnarray}
&& \mathrm{W}_{1,1}=\left(\begin{array}{ccc}
a_{+} & 0 & 0 \\
0 & b_{+} & 0 \\
0 & 0 & f_{+} \\ 
\end{array}
\right),~~
\mathrm{W}_{2,2}=\left(\begin{array}{ccc}
\bar{b}_{+} & 0 & 0 \\
0 & g & 0 \\
0 & 0 & \bar{b}_{-} \\ 
\end{array}
\right),~~
\mathrm{W}_{3,3}=\left(\begin{array}{ccc}
f_{-} & 0 & 0 \\
0 & b_{-} & 0 \\
0 & 0 & a_{-} \\ 
\end{array}
\right), \nonumber \\ 
&& \mathrm{W}_{1,2}=\left(\begin{array}{ccc}
0~ & 0~ & 0~ \\
c_{+} & 0~ & 0~ \\
0~ & d_{+} & 0~ \\ 
\end{array}
\right),~~
\mathrm{W}_{1,3}=\left(\begin{array}{ccc}
0~ & 0~ & 0~ \\
0~ & 0~ & 0~ \\
h~ & 0~ & 0~ \\ 
\end{array}
\right),~~
\mathrm{W}_{2,3}=\left(\begin{array}{ccc}
0~ & 0~ & 0~ \\
d_{-}~ & 0~ & 0~ \\
0~ & c_{-}~ & 0~ \\ 
\end{array}
\right),
\label{LAX1}
\end{eqnarray}
where the other matrix elements are determined by the time reversal symmetry (\ref{TIME}).

The above notation for the transition operator
bears the charge conjugation operation which exchange
the subscripts $+ \leftrightarrow -$ of the weights. We shall
see for instance that in the Hamiltonian limit
the asymmetry among the weights $c_{+}$ and $c_{-}$ implies the presence 
of a non-null azimuthal magnetic field.
This choice will be also convenient to
write compact expressions for the functional equations coming
from the Yang-Baxter equation.
We now substitute 
the matrix expression of the transition operator
in the Yang-Baxter equation using as an ansatz for $\mathrm{R}(\bf{w})$ 
the most general $9 \times 9$ matrix. The analysis of the corresponding
functional relations derived from Eq.(\ref{YBA}) reveals us 
that for a generic vector $\omega \in \mathbb{CP}^{13}$
the non null entries of the $\mathrm{R}$-matrix are also constrained by
the ice rule. This leads us to conclude that the basic form 
for the $\mathrm{R}(\bf{w})$-matrix is similar
to that of the transition operator, namely 
\EQ
\mathrm{R}(\bf{w})=\left[
\begin{array}{ccccccccc}
 \bf{a}_{+} & 0 & 0 & 0 & 0 & 0 & 0 & 0 & 0 \\
 0 & \bf{b}_{+} & 0 & \bf{c}_{+} & 0 & 0 & 0 & 0 & 0 \\
 0 & 0 & \bf{f}_{+} & 0 & \bf{d}_{+} & 0 & \bf{h} & 0 & 0 \\ 
 0 & {\bf{c}}_{+} & 0 & \overline{\bf{b}}_{+} & 0 & 0 & 0 & 0 & 0 \\
 0 & 0 & {\bf{d}}_{+} & 0 & \bf{g} & 0 & {\bf{d}}_{-} & 0 & 0 \\
 0 & 0 & 0 & 0 & 0 & \overline{\bf{b}}_{-} & 0 & {\bf{c}}_{-} & 0 \\
 0 & 0 & \bf{h} & 0 & \bf{d}_{-} & 0 & \bf{f}_{-} & 0 & 0 \\
 0 & 0 & 0 & 0 & 0 & \bf{c}_{-} & 0 & \bf{b}_{-} & 0 \\
 0 & 0 & 0 & 0 & 0 & 0 & 0 & 0 & \bf{a}_{-} \\
\end{array}
\right],
\label{RMA}
\EN
where we have distinguished the $\mathrm{R}$ elements by means of bold letters. 

In next section we shall investigate the solutions of the functional equations
derived by substituting the expressions of the transition operator (\ref{LAX1}) and
the $\mathrm{R}$-matrix (\ref{RMA}) into the Yang-Baxter equation. This leads to fifty-seven
distinct polynomial relations but fortunately only a subset of them are enough
to decide on the existence of a hypersurface solution. 

\section{The Functional Relations}
\label{sectionYBE}

The polynomial equations can be classified in terms of their
number of monomials involving the $\mathrm{R}$-matrix entries 
and two different set of Boltzmann weights. 
The minimum number of such monomials is 
three while the maximum one turns out
to be five and this information has been summarized in Table (\ref{TAB}).
\begin{table}[ht]
\begin{center}
\begin{tabular}{|c|c|}
\hline
Number of Relations & Number of Monomials \\ \hline
18 & Three \\ \hline
27 & Four \\ \hline
12 & Five \\ \hline
\end{tabular}
\caption{The number of functional relations with a given number 
monomials.} \label{TAB}
\end{center}
\end{table}

The first step to solve the Yang-Baxter relation is 
to perform the elimination of the $\mathrm{R}$-matrix elements. Because we
are dealing with homogeneous system of equations this
is equivalent of the vanishing of a number of determinants whose
coefficients depend on the Boltzmann weights. The main point is
to choose suitable system of relations whose determinants could
be factorized in the convenient form (\ref{fact}). We shall
start this analysis considering the simplest family of 
functional equations which are those involving
three different monomials.

\subsection{Three Monomials }
\label{subthree}

The expressions of the eighteen functional relations involving three different monomials are, 
\begin{eqnarray}
\label{threeq1}
&& {\bf{a}_{\pm}} {c}_{\pm}^{'} {a}_{\pm}^{''} - \overline{\bf{b}}_{\pm} {c}_{\pm}^{'} {b}_{\pm}^{''} - {\bf{c}_{\pm}} {a}_{\pm}^{'} {c}_{\pm}^{''}=0, \\
\label{threeq2}
&& {\bf{c}_{\pm}} {c}_{\pm}^{'} {b}_{\pm}^{''} + {\bf{b}_{\pm}} {a}_{\pm}^{'} {c}_{\pm}^{''} - {\bf{a}_{\pm}} {b}_{\pm}^{'} {c}_{\pm}^{''}=0, \\
\label{threeq3}
&& {\bf{a}_{\pm}} {d}_{\pm}^{'} {b}_{\pm}^{''} - {\bf{c}_{\pm}} {b}_{\pm}^{'} {d}_{\pm}^{''} - \overline{\bf{b}}_{\pm} {d}_{\pm}^{'} {f}_{\pm}^{''}=0, \\
\label{threeq4}
&& {\bf{b}_{\pm}} {b}_{\pm}^{'} {d}_{\pm}^{''} - {\bf{a}_{\pm}} {f}_{\pm}^{'} {d}_{\pm}^{''} + {\bf{c}_{\pm}} {d}_{\pm}^{'} {f}_{\pm}^{''}=0, \\
\label{threeq5}
&& {\bf{c}_{\pm}} {\bar{b}}_{\pm}^{'} {a}_{\pm}^{''} - \overline{\bf{b}}_{\pm} {c}_{\pm}^{'} {c}_{\pm}^{''} - {\bf{c}_{\pm}} {a}_{\pm}^{'} {\bar{b}}_{\pm}^{''}=0, \\
\label{threeq6}
&& {\bf{b}_{\pm}} {d}_{\pm}^{'} {d}_{\pm}^{''} + {\bf{c}_{\pm}} {\bar{b}}_{\mp}^{'} {f}_{\pm}^{''} - {\bf{c}_{\pm}} {f}_{\pm}^{'} {\bar{b}}_{\mp}^{''}=0, \\
\label{threeq7}
&& {\bf{f}_{\pm}} {b}_{\pm}^{'} {c}_{\mp}^{''} - {\bf{b}_{\pm}} {f}_{\pm}^{'} {c}_{\mp}^{''} + {\bf{d}_{\pm}} {d}_{\pm}^{'} {\bar{b}}_{\mp}^{''}=0, \\
\label{threeq8}
&& {\bf{d}_{\pm}} {f}_{\pm}^{'} {a}_{\mp}^{''} - {\bf{f}_{\pm}} {d}_{\pm}^{'} {c}_{\mp}^{''} - {\bf{d}_{\pm}} {\bar{b}}_{\mp}^{'} {\bar{b}}_{\mp}^{''}=0, \\
\label{threeq9}
&& \overline{\bf{b}}_{\pm} {d}_{\mp}^{'} {a}_{\pm}^{''} - {\bf{f}_{\mp}} {d}_{\mp}^{'} {b}_{\pm}^{''} - {\bf{d}_{\mp}} {\bar{b}}_{\pm}^{'} {c}_{\pm}^{''}=0.
\end{eqnarray}

We emphasize that each of the equations (\ref{threeq1}-\ref{threeq9})
splits into two different functional relations associated to the
two possible subscripts values $\pm$ for the weights. We start the solution
of these equations by first noticing that out of Eqs.(\ref{threeq1}-\ref{threeq6})
we can construct two independent homogeneous linear systems 
for the $\mathrm{R}$-matrix
entries ${\bf{a}}_{\pm}$,~${\bf{b}}_{\pm}$,~$\overline{\bf{b}}_{\pm}$ and ${\bf{c}}_{\pm}$.
We find that the determinants of coefficients associated to Eqs.(\ref{threeq1}-\ref{threeq4}) 
have the nice property that they can be written in the following factorized form, 
\EQ
\left[ ({b_{\pm}^{'}}^2-a_{\pm}^{'} f_{\pm}^{'})c_{\pm}^{''} d_{\pm}^{''}-({b_{\pm}^{''}}^2-a_{\pm}^{''} f_{\pm}^{''})c_{\pm}^{'} d_{\pm}^{'} \right] 
\left[ a_{\pm}^{'} d_{\pm}^{'} c_{\pm}^{''} f_{\pm}^{''} -b_{\pm}^{''} d_{\pm}^{''} b_{\pm}^{'} c_{\pm}^{'}\right]. 
\label{DET}
\EN

In order to have a non-trivial solution 
for the $\mathrm{R}$-matrix entries 
${\bf{a}}_{\pm}$,~${\bf{b}}_{\pm}$,~$\overline{\bf{b}}_{\pm}$ 
and ${\bf{c}}_{\pm}$ the above determinants
must vanish. This can be achieved if either of the 
two factors of Eq.(\ref{DET}) vanishes
which gives us two branches to be analyzed. 
We stress that the purpose of this paper
is not to pursue a classification of possible integrable 
nineteen vertex models even within the subclass of systems invariant
by the time reversal symmetry. Here we are mainly interested to point out
an example of solvable nineteen vertex model whose weights would be lying
on high-dimensional algebraic varieties.
In this sense we choose
the branch associated to the first factor of Eq.(\ref{DET}) 
since its polynomial expression is clearly less restrictive than that
of the second factor. We also note that the former factor  
has the suitable polynomial form (\ref{fact}) and the condition
that it vanishes leads us to our first divisor, 
\EQ
\frac{{b^{2}_{\pm}}-a_{\pm} f_{\pm}}{c_{\pm} d_{\pm}}= \Lambda_1^{\pm}.
\label{INV1}
\EN

We now can solve Eqs.(\ref{threeq1}-\ref{threeq4})
for the $\mathrm{R}$-matrix  
entries by means of linear elimination of such variables. 
After using Eq.(\ref{INV1}) and taking ${\bf{c}}_{\pm}$ 
as a common normalization we find 
that their expressions are,
\begin{eqnarray}
\label{RweightA}
&& \frac{{\bf{a}}_{\pm}}{{\bf{c}}_{\pm}}=\frac{b_{\pm}^{'} c_{\pm}^{'} b_{\pm}^{''} d_{\pm}^{''}-a_{\pm}^{'} d_{\pm}^{'} c_{\pm}^{''} f_{\pm}^{''}}{\Lambda_1^{\pm} c_{\pm}^{'} d_{\pm}^{'} c_{\pm}^{''} d_{\pm}^{''}},\\ \nonumber \\
\label{RweightBAR}
&& \frac{\overline{\bf{b}}_{\pm}}{{\bf{c}}_{\pm}}=
\frac{b_{\pm}^{'} c_{\pm}^{'} a_{\pm}^{''} d_{\pm}^{''}-a_{\pm}^{'} d_{\pm}^{'} b_{\pm}^{''} c_{\pm}^{''}}{\Lambda_1^{\pm} c_{\pm}^{'} d_{\pm}^{'} c_{\pm}^{''} d_{\pm}^{''}}, \\ \nonumber \\
\label{RweightB}
&& \frac{{\bf{b}}_{\pm}}{{\bf{c}}_{\pm}}=
\frac{f_{\pm}^{'} c_{\pm}^{'} b_{\pm}^{''} d_{\pm}^{''}-b_{\pm}^{'} d_{\pm}^{'} f_{\pm}^{''} c_{\pm}^{''}}{\Lambda_1^{\pm} c_{\pm}^{'} d_{\pm}^{'} c_{\pm}^{''} d_{\pm}^{''}},
\end{eqnarray}
where we are tacitly assuming that the free parameters $\Lambda_1^{\pm}$ are non null. Along the lines of our work \cite{MAR}
it is possible to show that the particular point $\Lambda_1^{\pm}=0$ corresponds to the branch  in which the second
factor of the determinant (\ref{DET}) is set to zero. As remarked before we shall not consider such rather special
branch in what follows.  

In order to complete the solution of the first twelve
functional relations we have to substitute 
these results for $\mathrm{R}$-matrix entries into the 
remaining equations (\ref{threeq5},\ref{threeq6}).
After few simplifications with the help of the 
divisor (\ref{INV1}) we find that the 
Eqs.(\ref{threeq5}) can be rewritten as,
\EQ
(b_{\pm}^{'} c_{\pm}^{'} -\Lambda_1^{\pm} {\bar{b}}_{\pm}^{'} d_{\pm}^{'})  a_{\pm}^{''} d_{\pm}^{''} -
(b_{\pm}^{''} c_{\pm}^{''} -\Lambda_1^{\pm} {\bar{b}}_{\pm}^{''} d_{\pm}^{''}) a_{\pm}^{'} d_{\pm}^{'} =0,
\EN
while Eqs.(\ref{threeq6}) becomes proportional to the polynomial,
\EQ
(b_{\pm}^{'} d_{\pm}^{'} -\Lambda_1^{\pm} c_{\pm}^{'} {\bar{b}}_{\mp}^{'} )  c_{\pm}^{''} f_{\pm}^{''} -
(b_{\pm}^{''} d_{\pm}^{''} -\Lambda_1^{\pm} c_{\pm}^{''} {\bar{b}}_{\pm}^{''} ) c_{\pm}^{'} f_{\pm}^{'} =0.
\EN

The above polynomials are both in the convenient 
factorized form (\ref{fact}) and they are solved by the 
following divisors, 
\EQ
\frac{b_{\pm} c_{\pm}-\Lambda_1^{\pm} {\bar{b}}_{\pm} d_{\pm}}{a_{\pm} 
d_{\pm}}= \Lambda_2^{\pm},
\label{INV2}
\EN
and 
\EQ
\frac{b_{\pm} d_{\pm}-\Lambda_1^{\pm} {\bar{b}}_{\mp} c_{\pm}}{c_{\pm} 
f_{\pm}}= \Lambda_3^{\pm}.
\label{INV3}
\EN

From the above analysis we are able to conclude that the number 
of independent weights
can be reduced by five variables. In fact, we first note that  
the divisors (\ref{INV1},\ref{INV2}) can be easily 
resolved by means of linear elimination of
the weights $d_{\pm}$ and $f_{\pm}$ and as result we obtain,
\begin{eqnarray}
\label{weightDMP}
&& d_{\pm}= \frac{b_{\pm}c_{\pm}}{\Lambda_2^{\pm}a_{\pm}+\Lambda_1^{\pm}{\bar{b}}_{\pm}}, \\ \nonumber \\
\label{weightFMP}
&& f_{\pm}= \frac{b_{\pm}}{a_{\pm}} \left[ \frac{ 
\Lambda_2^{\pm}a_{\pm}b_{\pm}+\Lambda_1^{\pm}(b_{\pm} {\bar{b}}_{\pm} -c^2_{\pm})}{
\Lambda_2^{\pm}a_{\pm}+\Lambda_1^{\pm}{\bar{b}}_{\pm}} \right].  
\end{eqnarray}

We next substitute the above variables into the last divisor (\ref{INV3}) and
notice that the weights $\bar{b}_{\pm}$ are still linearly encoded 
in the resulting expressions. This means that we can for instance
extract the weight $\bar{b}_{-}$ from 
Eq.(\ref{INV3}) by using the channel with subscript plus. Substituting 
this result
back to Eq.(\ref{INV3}) but now with subscript minus produces however a 
non-linear constraint among the variables $a_{\pm},b_{\pm},c_{\pm}$
and $\bar{b}_{+}$. Considering these steps we find that 
the expression for the weight $\bar{b}_{-}$ is,
\EQ
\label{weightBMbar}
{\bar{b}}_{-}= \frac{b_{+}}{\Lambda_1^{+} a_{+}} \left[ \frac{
(1-\Lambda_2^{+}\Lambda_3^{+})a_{+}b_{+}-\Lambda_1^{+} \Lambda_3^{+}(b_{+} {\bar{b}}_{+} -c^2_{+})}{ 
\Lambda_2^{+}a_{+}+\Lambda_1^{+}{\bar{b}}_{+} } \right],   
\EN
while the constraint turns out to be a degree five
hypersurface leaving on $\mathbb{CP}^{6}$ given by,
\begin{eqnarray}
\label{HYPER}
S[a_{\pm},b_{\pm},{\bar b}_{+},c_{\pm}]&=&[\Lambda_1^{+}]^2\left([1-\Lambda_2^{-} \Lambda_3^{-}] a_{-} b_{-}^2+\Lambda_1^{-} \Lambda_3^{-} b_{-} c_{-}^2-\Lambda_1^{-} \Lambda_2^{-} a_{-}^2 {\bar{b}}_{+} \right) a_{+} {\bar{b}}_{+} \nonumber \\
&+&\Lambda_1^{+} \Lambda_2^{+} \left( [1-\Lambda_2^{-} \Lambda_3^{-}] a_{-} b_{-}^2+\Lambda_1^{-} \Lambda_3^{-} b_{-} c_{-}^2-\Lambda_1^{-} \Lambda_2^{-} a_{-}^2 {\bar{b}}_{+} \right) a_{+}^2 \nonumber \\
&+& \Lambda_1^{+} \Lambda_1^{-} \Lambda_3^{+} \left(\Lambda_3^{-} b_{-}^2+\Lambda_1^{-} a_{-} {\bar{b}}_{+}\right) \left(b_{+} {\bar{b}}_{+}-c_{+}^2\right) b_{+} \nonumber \\
&+&\Lambda_1^{-} [\Lambda_2^{+} \Lambda_3^{+}-1] \left(\Lambda_3^{-} b_{-}^2+\Lambda_1^{-} a_{-} {\bar{b}}_{+}\right) a_{+} b_{+}^2=0.
\end{eqnarray}

Let us come back to discuss the solution of the 
remaining functional relations (\ref{threeq7}-\ref{threeq9}). Direct inspection
of such equations tells us that we 
can use two of them to eliminate the $\mathrm{R}$-matrix
entries ${\bf{d}}_{\pm}$ and ${\bf{f}}_{\pm}$. Choosing Eqs.(\ref{threeq7},\ref{threeq8})
to be solved we find that such matrix elements are given by,    
\begin{eqnarray}
\label{RweightD}
&& \frac{{\bf{d}}_{\pm}}{{\bf{c}}_{\pm}}= \frac{{\bf{b}}_{\pm}}{{\bf{c}}_{\pm}}
\left[ \frac{d_{\pm}^{'} f_{\pm}^{'} c_{\mp}^{''}}{b_{\pm}^{'} f_{\pm}^{'} a_{\mp}^{''} + [d_{\pm}^{'}]^2 {\bar b}_{\mp}^{''} - 
      b_{\pm}^{'} {\bar b}_{\mp}^{'} {\bar b}_{\mp}^{''}} \right] , \\  \nonumber \\
&& \frac{{\bf{f}}_{\pm}}{{\bf{c}}_{\pm}}= \frac{{\bf{d}}_{\pm}}{{\bf{c}}_{\pm}} \left[
\label{RweightF}
\frac{f_{\pm}^{'} a_{\mp}^{''} - {\bar b}_{\mp}^{'} {\bar b}_{\mp}^{''}}{d_{\pm}^{'} c_{\mp}^{''} }\right ],
\end{eqnarray}
where the ratio ${\bf{b}}_{\pm}/{\bf{c}}_{\pm}$ is obtained from Eq.(\ref{RweightB}).

We have now reached a point in which only Eqs.(\ref{threeq9}) remain to be analyzed.
They lead to polynomials having as unknowns the $\mathrm{R}$-matrix entries ${\bf{c}}_{\pm}$
whose coefficients consist of very complicated expressions depending on the weights
$a_{\pm},b_{\pm},c_{\pm}$ and $\bar{b}_{+}$. We shall postpone their analysis until
next section since many of the free parameters $\Lambda_1^{\pm}$, $\Lambda_2^{\pm}$
and $\Lambda_3^{\pm}$ are going to be fixed later on. This fact will be 
responsible for the cancellation
a large number of monomials resulting in much simpler polynomial expressions. 
In spite of that it is possible to make a prediction on the lowest dimension value
of the variety $\mathrm{Y}$ associated to a potential solution of the Yang-Baxter equation.
First we should note 
that Eqs.(\ref{threeq9}) are only capable to produce at most 
two more additional divisors. This fact together
with the results obtained so far imply that
the maximum number of divisors solving 
the full set of functional 
equations (\ref{threeq1}-\ref{threeq9}) 
should be therefore eight. We next recall that these divisors are embedded in a $\mathbb{CP}^{11}$ 
projective space whose coordinates are the twelve 
weights $a_{\pm},b_{\pm},{\bar{b}}_{\pm},c_{\pm},d_{\pm}$ and $f_{\pm}$. 
Considering
these information on formula (\ref{DIM}) we conclude that the dimension of the
underlying variety must satisfy,
\EQ
\mathrm{dim}(\mathrm{Y}) \geq 12-1-8=3,
\label{bound}
\EN
and therefore we have a concrete possibility of the Boltzmann weights being sited 
at least on an algebraic threefold.

The above conclusion assumes that the many other functional relations
coming from the Yang-Baxter equation can be solved
without additional divisors other than that necessary to resolve
Eqs.(\ref{threeq9}) as well as those that determine the last two weights $h$ and $g$.
It is exactly at this point that
the free parameters $\Lambda_1^{\pm}, \Lambda_2^{\pm}$ and $\Lambda_3^{\pm}$ 
we have at our disposal are 
going to play a very important role.
In fact, this freedom will be used to cancel out several
relevant polynomial relations avoiding 
the need of the proposal 
of any extra divisors.

\subsection{Four Monomials }

The weights $h$ and $g$ start to emerge on the functional relations containing four
different monomials.  The key point to solve these relations is to notice that few
of them  still have as unknowns some $\mathrm{R}$-matrix entries that have already
been sorted out in subsection \ref{subthree}. The condition of consistence 
of our elimination procedure compels us
to search for linear combinations among such relations and four 
special functional equations
involving three monomials. The vanishing of the determinants of these linear combinations
lead us to determine the remaining weights $h$ and $g$ with the help of only two
independent divisors. This makes it possible to maintain the lower bound (\ref{bound}) 
for the dimension of the underlying algebraic variety. Let us 
first describe this reasoning for the weight
$h$.

\subsubsection{ The variables $h$ and ${\bf{h}}$ }

Among the several four monomials functional relations solely four of them have
as unknowns the $\mathrm{R}$-matrix elements we have previously eliminated whose
coefficients also contain the weights $h^{'}$ and $h^{''}$. These 
relations involve the entries ${\bf{b}}_{\pm},\overline{\bf{b}}_{\pm}$ and 
${\bf{c}}_{\pm}$ and their expressions are,
\begin{eqnarray}
\label{linearfoureq1}
&& {\bf{c}}_{+} d_{+}^{'} d_{-}^{''} - {\bf{c}}_{-} d_{-}^{'} d_{+}^{''} + {\overline{\bf{b}}}_{+} {\bar{b}}_{-}^{'} h^{''} - 
     {\overline{\bf{b}}}_{-} {\bar{b}}_{+}^{'} h^{''}=0,\\
\label{linearfoureq2}
&& {\overline{\bf{b}}}_{-} c_{+}^{'} c_{-}^{''} - {\bf{b}}_{+} d_{+}^{'} d_{-}^{''} - {\bf{c}}_{+} {\bar{b}}_{-}^{'} h^{''} + {\bf{c}}_{-} h^{'} {\bar{b}}_{-}^{''}=0, \\
\label{linearfoureq3}
&& {\overline{\bf{b}}}_{+} c_{-}^{'} c_{+}^{''} - {\bf{b}}_{-} d_{-}^{'} d_{+}^{''} - {\bf{c}}_{-} {\bar{b}}_{+}^{'} h^{''} + 
     {\bf{c}}_{+} h^{'} {\bar{b}}_{+}^{''}=0, \\
\label{linearfoureq4}
&& {\bf{c}}_{-} c_{+}^{'} c_{-}^{''} - {\bf{c}}_{+} c_{-}^{'} c_{+}^{''} + {\bf{b}}_{-} h^{'} {\bar{b}}_{-}^{''} - {\bf{b}}_{+} h^{'} {\bar{b}}_{+}^{''}=0.
\end{eqnarray}

In order to build out a consistent homogeneous linear system 
we have to search for two more relations involving
the  six matrix elements
${\bf{b}}_{\pm},\overline{\bf{b}}_{\pm}$ and 
${\bf{c}}_{\pm}$. Direct inspection of the functional equations 
with three monomials reveals us that
such relations are in fact given by the four possible channels of Eqs.(\ref{threeq5},\ref{threeq6}). 
The composition of the eight functional relations 
(\ref{threeq5},\ref{threeq6},\ref{linearfoureq1}-\ref{linearfoureq4}) gives rise
to twenty-eight different linear systems 
but we find that only two of them are able to produce
determinants that factorize into smaller pieces. This turns out to be
the combination of the above relations (\ref{linearfoureq1}-\ref{linearfoureq4}) with
either the plus or the minus subscript  component
of the three monomials equations (\ref{threeq5},\ref{threeq6}). The vanishing
of such determinants can be written as,
\EQ
\left[c_{\pm}^{'} d_{\mp}^{'} c_{\mp}^{''} d_{\pm}^{''} - h^{'} {\bar{b}}_{\pm}^{'} h^{''} {\bar{b}}_{\mp}^{''} \right]
F_{\pm}(\omega^{'},\omega^{''})=0,
\label{DETH}
\EN
where the polynomial 
$F_{\pm}(\omega^{'},\omega^{''})$ has the following more complicated expression,
\begin{eqnarray}
F_{\pm}(\omega^{'},\omega^{''}) &=& 
c_{\mp}^{'} d_{\pm}^{'} {\bar{b}}_{\pm}^{'} h^{'} a_{\pm}^{''} c_{\pm}^{''} d_{\pm}^{''} {\bar{b}}_{\mp}^{''}-a_{\pm}^{'} c_{\pm}^{'} d_{\pm}^{'} {\bar{b}}_{\mp}^{'} c_{\mp}^{''} d_{\pm}^{''} {\bar{b}}_{\pm}^{''} h^{''}
+c_{\pm}^{'} d_{\mp}^{'} {\bar{b}}_{\mp}^{'} h^{'} c_{\pm}^{''} d_{\pm}^{''} f_{\pm}^{''} {\bar{b}}_{\pm}^{''} \nonumber \\
&-&c_{\pm}^{'} d_{\pm}^{'} f_{\pm}^{'} {\bar{b}}_{\pm}^{'} c_{\pm}^{''} d_{\mp}^{''} {\bar{b}}_{\mp}^{''} h^{''} 
+[c_{\pm}^{'} d_{\pm}^{'}]^2 c_{\mp}^{''} c_{\pm}^{''} d_{\mp}^{''} d_{\pm}^{''} - c_{\mp}^{'} c_{\pm}^{'} d_{\mp}^{'} d_{\pm}^{'} [c_{\pm}^{''} d_{\pm}^{''}]^2 \nonumber \\
&+&c_{\pm}^{'} d_{\pm}^{'} {\bar{b}}_{\pm}^{'} {\bar{b}}_{\mp}^{'} \left( a_{\pm}^{''} c_{\mp}^{''} d_{\pm}^{''} h^{''}+c_{\pm}^{''} d_{\mp}^{''} f_{\pm}^{''} h^{''}-c_{\pm}^{''} d_{\pm}^{''} [h^{''}]^2 \right) \nonumber \\
&-&c_{\pm}^{''} d_{\pm}^{''} {\bar{b}}_{\pm}^{''} {\bar{b}}_{\mp}^{''} \left(a_{\pm}^{'} c_{\mp}^{'} d_{\pm}^{'} h^{'}+c_{\pm}^{'} d_{\mp}^{'} f_{\pm}^{'} h^{'}-c_{\pm}^{'} d_{\pm}^{'}[h^{'}]^2 \right).
\end{eqnarray}

We encounter once again the situation in which we have 
in principle two branches that have to be
analyzed depending on the factor 
in Eq.(\ref{DETH}) we set to zero. As already emphasized this work is not
concerned with the complete classification of nineteen vertex models and 
in what follows we will choose the branch in which the
weights $h^{'}$ and $h^{''}$ can
be determined in a linear manner. 
This leads us to impose
the vanishing of the first term in Eq.(\ref{DETH}) since the polynomials 
$F_{\pm}(\omega^{'},\omega^{''})$ contain clearly quadratic terms 
on the weights $h^{'}$ and $h^{''}$. The corresponding divisors in which 
the single and double primed variables are separated from each other are,
\EQ
\frac{c_{+}^{'} d_{-}^{'}}{h^{'} {\bar{b}}_{+}^{'}}=
\frac{h^{''} {\bar{b}}_{-}^{''}}{c_{-}^{''}d_{+}^{''}}= \Lambda_4^{+}~~\mathrm{and}~~
\frac{c_{+}^{''} d_{-}^{''}}{h^{''} {\bar{b}}_{+}^{''}}=
\frac{h^{'} {\bar{b}}_{-}^{'}}{c_{-}^{'}d_{+}^{'}}= \Lambda_4^{-}.
\label{INV4pm}
\EN

We now have the necessary condition to start the solution of the functional relations
(\ref{linearfoureq1}-\ref{linearfoureq4}). The first step is to extract the ratio
among the $\mathrm{R}$-matrix elements ${\bf{c}}_{+}$ and ${\bf{c}}_{-}$ from one
of these equations. We find that Eq.(\ref{linearfoureq1}) gives us the
simplest possible  expression for such ratio which is,
\EQ
\frac{ {\bf{c}}_{-} }{ {\bf{c}}_{+} }= \frac{c_{-}^{'} {\bar{b}}_{-}^{'} c_{-}^{''} }{ c_{+}^{'} {\bar{b}}_{+}^{'} c_{+}^{''}}
\left[ \frac{ \Lambda_4^{-} b_{+}^{'} [c_{+}^{'}]^2 {\bar{b}}_{+}^{''} + 
       {\bar{b}}_{-}^{'} (\Lambda_2^{+} a_{+}^{'} + \Lambda_1^{+} {\bar{b}}_{+}^{'}) ({\bar{b}}_{+}^{'} a_{+}^{''} - a_{+}^{'} {\bar{b}}_{+}^{''})} {
     \Lambda_4^{-} b_{+}^{'} [c_{-}^{'}]^2 {\bar{b}}_{-}^{''} + 
       {\bar{b}}_{-}^{'} (\Lambda_2^{+} a_{+}^{'} + \Lambda_1^{+} {\bar{b}}_{+}^{'}) ({\bar{b}}_{-}^{'} a_{-}^{''} - a_{-}^{'} {\bar{b}}_{-}^{''})} \right].
\label{ratiocpcm}
\EN

We next substitute this result in Eq.(\ref{linearfoureq2}) which leads us to a 
degree three bihomogenous polynomial 
on the weights $a_{+},b_{+},\bar{b}_{+}$ and $c_{+}$. We find that such polynomial
satisfies the anti-symmetrical property (\ref{ANTI}) only when 
the free parameters $\Lambda_4^{\pm}$ are related by, 
\EQ
\Lambda_4^{-}=\Lambda_4^{+},
\label{cons1}
\EN
bringing the divisors (\ref{INV4pm}) to have the same form 
on both single and double primed indices as expected. These divisors
can now be used in order to determine the weights $h$ and $b_{-}$ 
in terms of the so far free amplitudes
$a_{+},b_{+},\bar{b}_{+}$ and $c_{+}$. By substituting 
Eqs.(\ref{weightDMP},\ref{weightBMbar}) in Eqs.(\ref{INV4pm})
and after few simplifications their expressions become,
\begin{eqnarray}
\label{weightH}
&& h= \frac{\Lambda_1^{+} \Lambda_4^{+} a_{+} c_{+} c_{-}}
{[1-\Lambda_2^{+} \Lambda_3^{+}] a_{+} b_{+}
+ \Lambda_1^{+} \Lambda_3^{+}[c_{+}^2-b_{+} \bar{b}_{+}]}, \\ \nonumber \\
\label{weightBM}
&&b_{-}=\frac{
\Lambda_1^{-}[1-\Lambda_2^{+} \Lambda_3^{+}]a_{+}b_{+}^2 +\Lambda_1^{+} \Lambda_2^{-}[
\Lambda_2^{+}a_{+}+\Lambda_1^{+}\bar{b}_{+}]a_{-}a_{+} +\Lambda_1^{+}\Lambda_1^{-}\Lambda_3^{+}
[c_{+}^2-b_{+} \bar{b}_{+}]b_{+}}
{[1-\Lambda_2^{+} \Lambda_3^{+}] a_{+} b_{+}
+ \Lambda_1^{+} \Lambda_3^{+}[c_{+}^2-b_{+} \bar{b}_{+}]} \nonumber \\
&&~~~~\times \frac{[\Lambda_4^{+}]^2 \bar{b}_{+}}{\Lambda_2^{+}a_{+}+\Lambda_1^{+}\bar{b}_{+}}.
\end{eqnarray}

Let us now return to the analysis of Eq.(\ref{linearfoureq2}). We find that using the
constraint (\ref{cons1}) the vanishing of this functional equation becomes
equivalent to the following relation,
\EQ
\label{eqaux}
\bar{b}_{+}^{'} a_{+}^{''}\mathrm{H}(a_{+}^{'},b_{+}^{'},{\bar{b}}_{+}^{'}) 
\mathrm{G}(a_{+}^{''},b_{+}^{''},{\bar{b}}_{+}^{''},c_{+}^{''})- 
\bar{b}_{+}^{''} a_{+}^{'}\mathrm{H}(a_{+}^{''},b_{+}^{''},{\bar{b}}_{+}^{''}) 
\mathrm{G}(a_{+}^{'},b_{+}^{'},{\bar{b}}_{+}^{'},c_{+}^{'})=0, 
\EN
where the expressions for the polynomials 
$\mathrm{H}(a_{+},b_{+},{\bar{b}}_{+})$ 
and $\mathrm{G}(a_{+},b_{+},{\bar{b}}_{+},c_{+})$ are, 
\begin{eqnarray}
&& \mathrm{H}(a_{+},b_{+},{\bar{b}}_{+}) =
(1-\Lambda_2^{+}\Lambda_3^{+})a_{+}b_{+}
+\Lambda_1^{+}\Lambda_3^{+}(c^2_{+}-b_{+}{\bar{b}}_{+}), \\
&& \mathrm{G}(a_{+},b_{+},{\bar{b}}_{+},c_{+}) =
\Lambda_3^{+}(1-\Lambda_2^{+}\Lambda_3^{+})a_{+}b_{+} +\Lambda_1^{+}(\Lambda_3^{+}-\Lambda_4^{+})\left[
\Lambda_3^{+}c^2_{+}-(\Lambda_3^{+}+\Lambda_4^{+})b_{+} {\bar{b}}_{+} \right ].
\end{eqnarray}

We see that Eq.(\ref{eqaux}) has the desirable factorized form (\ref{fact}) and in principle
could be solved by means of an extra divisor between the weights 
$a_{+},b_{+},\bar{b}_{+}$ and $c_{+}$. This certainly is going to lower the bound of the underlying
algebraic variety and therefore such solution should be 
discarded whenever possible. It is however fortunate
that Eq.(\ref{eqaux}) can be trivially satisfied provide we impose the following
relations among the free parameters,
\EQ
\label{cons2}
\Lambda_4^{+}=\Lambda_3^{+}=\frac{1}{\Lambda_2^{+}},
\EN
which sets the polynomial
$\mathrm{G}(a_{+},b_{+},{\bar{b}}_{+},c_{+})$ identically to zero.

The above reasoning can also be implemented to 
Eqs.(\ref{linearfoureq3},\ref{linearfoureq4}) which
at this point are proportional to each other. The
basic difference is that now we have the presence of the weight $c_{-}$ in their
expressions however solely through even powers. This variable can then be eliminated
in a systematic way with the help of the 
hypersurface expression (\ref{HYPER}) since its dependence on the weight $c_{-}$ 
is quadratic. It is not difficult to perform this algebraic operation 
within the Mathematica computer system as exemplified in Appendix A.
After this operation we have an involved degree 
eight bihomogenous polynomial
which fortunately vanishes by imposing the additional constraints,
\EQ
\label{cons3}
\Lambda_3^{-}=\frac{1}{\Lambda_2^{-}}=\Lambda_2^{+}.
\EN

At this point we have now gathered the basic ingredients
to finally come back to the solution of the functional relations (\ref{threeq9}).
Considering the value for
the weight $b_{-}$ given by Eq.(\ref{weightBM}) as well as the constraints
(\ref{cons2},\ref{cons3}) we find that the resulting polynomials coming
from Eqs.(\ref{threeq9}) are magically canceled out provided that the
weight $a_{-}$ is fixed by the following expression,
\EQ
\label{weightAM}
a_{-}= \frac{(b_{+} \bar{b}_{+}-c_{+}^2)\left(\Lambda_1^{-}a_{+}b_{+}-
\Lambda_2^{+}[c_{+}^2-b_{+} \bar{b}_{+}]\right)}{a_{+}^2(\Lambda_2^{+}a_{+}+\Lambda_1^{+}\bar{b}_{+})}. 
\EN

In order to complete the solution we need to exhibit the $\mathrm{R}$-matrix
element $\bf{h}$. There are several functional equations with four
monomials encoding the variable $\bf{h}$ together with the previous
determined weights and $\mathrm{R}$-matrix entries. These relations are
listed below,
\begin{eqnarray}
\label{fourHeq1}
&& {\bf{h}} d_{\mp}^{'} b_{\pm}^{''} + {\bf{d}}_{\pm} {\bar{b}}_{\pm}^{'} c_{\pm}^{''} - {\bf{c}}_{\pm} b_{\pm}^{'} d_{\mp}^{''} - {\overline{\bf{b}}}_{\pm} d_{\pm}^{'} h^{''}=0, \\
&& {\bf{h}} h^{'} f_{\pm}^{''} +{\bf{d}}_{\pm} c_{\pm}^{'} d_{\pm}^{''} + {\bf{f}}_{\pm} a_{\pm}^{'} h^{''} 
- {\bf{a}}_{\pm} f_{\pm}^{'} h^{''}=0, \\
&& {\bf{h}} f_{\pm}^{'} a_{\mp}^{''} - {\bf{h}} a_{\mp}^{'} f_{\pm}^{''} 
- {\bf{f}}_{\pm} h^{'} h^{''}
- {\bf{d}}_{\pm} c_{\mp}^{'} d_{\pm}^{''} =0,\\
&& {\bf{h}} a_{\mp}^{'} h^{''} -{\bf{a}}_{\mp} h^{'} a_{\mp}^{''} + {\bf{d}}_{\pm} 
c_{\mp}^{'} d_{\mp}^{''} + {\bf{f}}_{\pm} h^{'} f_{\mp}^{''}=0,\\
&& {\bf{h}} b_{\pm}^{'} c_{\mp}^{''} - {\bf{b}}_{\pm} h^{'} c_{\pm}^{''} + {\bf{d}}_{\mp} d_{\pm}^{'} {\bar{b}}_{\mp}^{''}
-{\bf{c}}_{\pm} c_{\mp}^{'} b_{\pm}^{''}=0, \\ 
\label{fourHeq6}
&& {\bf{h}} b_{+}^{'} b_{-}^{''} - {\bf{h}} b_{-}^{'} b_{+}^{''} + {\bf{d}}_{-} d_{+}^{'} c_{-}^{''} - {\bf{d}}_{+} d_{-}^{'} c_{+}^{''}=0.
\end{eqnarray}

We start the solution of the above relations solving one of them for the ratio
${\bf{h}}/{\bf{c}}_{+}$. Here we choose the component of Eqs.(\ref{fourHeq1}) 
with subscript plus
and we find that this ratio is given by,
\begin{eqnarray}
\frac{{\bf{h}}}{{\bf{c}}_{+}} &=&
\left[\frac{[\Lambda_2^{+}]^2 (\Lambda_2^{+} a_{+}^{'}+\Lambda_1^{+} {\bar{b}}_{+}^{'}) \left[ \Lambda_1^{-} a_{+}^{'} b_{+}^{'}+\Lambda_2^{+} (b_{+}^{'} {\bar{b}}_{+}^{'}-[c_{+}^{'}]^2) \right] -a_{+}^{'} b_{+}^{'} {\bar{b}}_{+}^{'}}{ 
[\Lambda_2^{+}]^2 (\Lambda_2^{+} a_{+}^{'}+\Lambda_1^{+} {\bar{b}}_{+}^{'}) \left[\Lambda_1^{-} a_{+}^{''} b_{+}^{''}+\Lambda_2^{+} (b_{+}^{''} {\bar{b}}_{+}^{''}-[c_{+}^{''}]^2)\right] -a_{+}^{''} b_{+}^{''} {\bar{b}}_{+}^{'}} \right] \nonumber \\
&\times&
\left[\frac{
[a_{+}^{''}]^2 c_{-}^{''}
(b_{+}^{'} {\bar{b}}_{+}^{'}-[c_{+}^{'}]^2) (\Lambda_2^{+} a_{+}^{''} + \Lambda_1^{+} {\bar{b}}_{+}^{''})}{
[a_{+}^{'}]^2 c_{-}^{'}
(b_{+}^{''} {\bar{b}}_{+}^{''}-[c_{+}^{''}]^2) (\Lambda_2^{+} a_{+}^{'} + \Lambda_1^{+} {\bar{b}}_{+}^{'})} \right].
\end{eqnarray}

We now substitute the results we have obtained so far in the 
remaining relations (\ref{fourHeq1}-\ref{fourHeq6}) resulting
in polynomials depending on the weights $a_{+},b_{+},\bar{b}_{+},c_{+}$ and $c_{-}$.
As before the dependence on the variable $c_{-}$ occurs only via even powers and this
weight can once again be eliminated by using the hypersurface constraint (\ref{HYPER}). After performing this step
we find that all the above functional relations are satisfied at the expense of fixing
the parameter $\Lambda_1^{-}$ as,
\EQ
\label{LAM1M}
\Lambda_1^{-}=\frac{1-[\Lambda_2^{+}]^2+[\Lambda_2^{+}]^4}{\Lambda_1^{+} [\Lambda_2^{+}]^2}.
\EN

We conclude observing that so far we have not introduced any additional divisor 
besides the three expected ones. In fact, two of them are 
used to resolve of Eqs.(\ref{threeq9})
and are directly associated with the determination of the 
weights $a_{-}$ and $b_{-}$ by means of 
the free variables
$a_{+},b_{+},\bar{b}_{+}$ and $c_{+}$. The third one is responsible 
for the elimination of the 
weight $h$ again in terms
of the same variables.  This means that the general character
of our solution concerning the variety 
dimension bound (\ref{bound}) remains unchanged.

\subsubsection{The variables $g$ and ${\bf{g}}$}

We have six functional relations encoding the weights $g^{'}$, $g^{''}$ and the matrix element
${\bf{g}}$ together with the previous determined variables. Their expressions are given by,
\begin{eqnarray}
\label{fourGeq1}
&& {\bf{c}}_{\pm} g^{'} b_{\pm}^{''} + {\bf{b}}_{\pm} c_{\pm}^{'} c_{\pm}^{''} - {\overline{\bf{b}}}_{\pm} d_{\pm}^{'} d_{\pm}^{''} 
- {\bf{c}}_{\pm} b_{\pm}^{'} g^{''}=0, \\
\label{fourGeq2}
&& {\bf{g}} d_{\pm}^{'} {\bar{b}}_{\mp}^{''}
+{\bf{d}}_{\pm} b_{\pm}^{'} c_{\mp}^{''} - {\bf{c}}_{\pm} {\bar{b}}_{\mp}^{'} d_{\pm}^{''} - {\bf{b}}_{\pm} d_{\pm}^{'} g^{''}=0, \\ 
\label{fourGeq3}
&& {\bf{g}} {\bar{b}}_{\mp}^{'} c_{\mp}^{''} 
+{\bf{d}}_{\pm} d_{\pm}^{'} b_{\mp}^{''} - {\overline{\bf{b}}}_{\mp} g^{'} c_{\mp}^{''}
 -{\bf{c}}_{\mp} c_{\mp}^{'} {\bar{b}}_{\mp}^{''}=0.
\end{eqnarray}

We note that the unknowns of Eqs.(\ref{fourGeq1}) are again the $\mathrm{R}$-matrix elements 
${\bf{b}}_{\pm},\overline{\bf{b}}_{\pm}, 
{\bf{c}}_{\pm}$ and thus we have to
make linear combinations with the same three monomials functional relations. This leads us to set to zero
the determinants built out of either the plus or the minus 
channels of Eqs.(\ref{threeq5},\ref{threeq6},\ref{fourGeq1}).
After using the help of divisors (\ref{INV2},\ref{INV3}) these vanishing conditions become,
\EQ
b_{\pm}^{'}c_{\pm}^{'}d_{\pm}^{'} \left( \Lambda_1^{+}c_{\pm}^{''} d_{\pm}^{''} g^{''}+a_{\pm}^{''} [d_{\pm}^{''}]^2 
+[c_{\pm}^{''}]^2 f_{\pm}^{''} \right) 
-b_{\pm}^{''}c_{\pm}^{''}d_{\pm}^{''} \left( \Lambda_1^{+}c_{\pm}^{'} d_{\pm}^{'} g^{'}+a_{\pm}^{'} [d_{\pm}^{'}]^2 
+[c_{\pm}^{'}]^2 f_{\pm}^{'} \right)=0.
\EN

Clearly, the above determinants give rise to 
polynomials of the 
factorized form (\ref{fact}) and the corresponding
divisors are given by,
\EQ
\label{INV5}
\frac{\Lambda_1^{\pm}c_{\pm} d_{\pm} g+a_{\pm} d_{\pm}^2+c_{\pm}^2 f_{\pm}}{b_{\pm}c_{\pm}d_{\pm}}=\Lambda_5^{\pm}. 
\EN

Because we have two divisors to eliminate a single weight $g$ 
they must be compatible otherwise 
they will give origen to an extra constrain. This would be an 
undesirable situation since it
will expoil the earlier bound (\ref{bound}) for 
the variety dimension. Fortunately,  
this compatibility can be achieved once the
parameters $\Lambda_5^{\pm}$ are fixed by the relation,
\EQ
\Lambda_5^{+}=\Lambda_5^{-}= \frac{1+[\Lambda_2^{+}]^2}{\Lambda_2^{+}},
\EN
and by using Eqs.(\ref{weightDMP},\ref{weightFMP}) the weight $g$ is uniquely 
determined by the expression,
\EQ
\label{weightG}
g=\frac{a_{+}b_{+}\bar{b}_{+}+\Lambda_2^{+}(\Lambda_2^{+}a_{+}
+\Lambda_1^{+}\bar{b}_{+})(c_{+}^2-b_{+}\bar{b}_{+})}{\Lambda_2^{+}a_{+}(\Lambda_2^{+}+\Lambda_1^{+}\bar{b}_{+})}.
\EN

In order to complete the solution of the above functional 
relations we just need to solve one of them for the
matrix element ${\bf{g}}$. We can for instance extract this 
$\mathrm{R}$-matrix entry from the plus component of
Eq.(\ref{fourGeq2}) to obtain,
\EQ
\label{ratioG}
\frac{{\bf{g}}}{{\bf{c}}_{+}}= \left[ \frac{{\bf{b}}_{+}}{{\bf{c}}_{+}} d_{+}^{'} g^{''} - 
\frac{ {\bf{d}}_{+}}{ {\bf{c}}_{+}} b^{'}_{+} c_{-}^{''} -\bar{b}_{-}^{'} d_{+}^{''} \right ]/(d_{+}^{'} \bar{b}_{-}^{''}),
\EN
where the ratios 
${\bf{b}}_{+}/{\bf{c}}_{+}$ and ${\bf{d}}_{+}/{\bf{c}}_{+}$ are given
by Eqs.(\ref{RweightB},\ref{RweightD}). 

By substituting the result (\ref{ratioG}) back into 
Eqs.(\ref{fourGeq1}-\ref{fourGeq3}) we find
that they are either zero or produce polynomials depending on the variables 
$a_{+},b_{+},\bar{b}_{+}, c_{+}$ and $c_{-}$. In the latter case 
we use the hypersurface (\ref{HYPER}) to eliminate the even powers of the 
weight $c_{-}$ and after this procedure they
are immediately satisfied. 
At this point we mention that we still have to verify
some additional functional relations involving four 
and five monomials which so far have not been mentioned in the text.
These extra relations are presented in 
Appendix A in which we have discussed the 
algebraic procedure used to check
that they are indeed satisfied. 

As a result the solution of the Yang-Baxter equation is determined
by the intersection of the  
hypersurface (\ref{HYPER}) with the expressions of the weights 
$a_{-}$ and $b_{-}$ given by
Eqs.(\ref{weightBM},\ref{weightAM}). This is clearly a complete 
intersection leadings us to a three-dimensional variety 
whose polynomial can be written as,
\begin{eqnarray}
\label{TFOLD}
\mathrm{T}(a_{+},b_{+},\bar{b}_{+},c_{+},c_{-})&=&
(c_{+}^2-b_{+}\bar{b}_{+})^2 \Bigg[ 
\Lambda_2^{+}(c_{+}^2-b_{+}\bar{b}_{+})(\Lambda_2^{+}a_{+}+\Lambda_1^{+}\bar{b}_{+})
-\Lambda_{1}^{-}\Lambda_2^{+}a_{+}^2b_{+} \nonumber \\ 
&-& ([\Lambda_2^{+}]^2-1)a_{+}b_{+}\bar{b}_{+} \Bigg]
-a_{+}^3c_{-}^2\left[\Lambda_2^{+}a_{+} + \Lambda_1^{+}\bar{b}_{+}\right]^2, 
\end{eqnarray}
where $\Lambda_1^{+}$ and $\Lambda_2^{+}$ are free couplings while $\Lambda_1^{-}$ is
determined by these parameters using Eq.(\ref{LAM1M}).

The transition 
operator weights other than
those entering as variables in the above threefold 
can be determined with the help of the divisors (\ref{weightDMP},\ref{weightFMP},\ref{weightBMbar},\ref{weightH},
\ref{weightBM},\ref{weightAM},\ref{weightG}). For sake of clarify we list below their explicit 
expressions
\footnote{We observe the existence of the
simple relation $b_{+} b_{-}=\bar{b}_{+} \bar{b}_{-}$.},
\begin{eqnarray}
\label{BB1}
&&d_{+}=\frac{b_{+}c_{+}}{\Lambda_2^{+}a_{+}+\Lambda_1^{+}\bar{b}_{+}},~~
f_{+}=\frac{b_{+}\left(\Lambda_2^{+}a_{+}b_{+}-\Lambda_1^{+}[c_{+}^2-b_{+}\bar{b}_{+}]\right)}
{a_{+}[\Lambda_2^{+}a_{+}+\Lambda_1^{+}\bar{b}_{+}]}, \\ \nonumber \\
&&b_{-}=\frac{\bar{b}_{+}[c_{+}^2-b_{+}\bar{b}_{+}]}{\Lambda_2^{+}a_{+}[\Lambda_2^{+}a_{+}+\Lambda_1^{+}\bar{b}_{+}]},~~
\bar{b}_{-}=\frac{b_{+}[c_{+}^2-b_{+}\bar{b}_{+}]}{\Lambda_2^{+}a_{+}[\Lambda_2^{+}a_{+}+\Lambda_1^{+}\bar{b}_{+}]}, \\ \nonumber \\
&&g=\frac{a_{+}b_{+}\bar{b}_{+}+\Lambda_2^{+}[c_{+}^2-b_{+}\bar{b}_{+}][\Lambda_2^{+}a_{+}+\Lambda_1^{+}\bar{b}_{+}]}
{\Lambda_2^{+}a_{+}[\Lambda_2^{+}a_{+}+\Lambda_1^{+}\bar{b}_{+}]},~~
h=\frac{a_{+}c_{+}c_{-}}{c_{+}^2-b_{+}\bar{b}_{+}}, \\ \nonumber \\
&&a_{-}=\frac{[c_{+}^2-b_{+}\bar{b}_{+}]\left(\Lambda_2^{+}[c_{+}^2-b_{+}\bar{b}_{+}]-\Lambda_1^{-}a_{+}b_{+}\right)}
{a_{+}^2[\Lambda_2^{+}a_{+}+\Lambda_1^{+}\bar{b}_{+}]},~~ 
d_{-}=\frac{a_{+}\bar{b}_{+}c_{-}}{\Lambda_2^{+}[c_{+}^2-b_{+}\bar{b}_{+}]},
\\ \nonumber \\ 
\label{BB2}
&&f_{-}=\frac{\bar{b}_{+}\left(\Lambda_1^{-}\Lambda_2^{+}a_{+}^3c_{-}^2[\Lambda_2^{+}a_{+}+\Lambda_1^{+}\bar{b}_{+}]^2 
-\bar{b}_{+}[c_{+}^2-b_{+}\bar{b}_{+}]^3\right)}{[\Lambda_2^{+}]^2[\Lambda_2^{+}a_{+}+\Lambda_1^{+}\bar{b}_{+}]
[c_{+}^2-b_{+}\bar{b}_{+}]^2\left(\Lambda_1^{-}a_{+}b_{+}-\Lambda_2^{+}[c_{+}^2-b_{+}\bar{b}_{+}]\right)}.
\end{eqnarray}

The explicit form of the $\mathrm{R}$-matrix as function of the weights with
distinct labels are in general very cumbersome. In order to avoid overcrowding this section with
extra heavier formulae we have presented the 
$\mathrm{R}$-matrix elements in Appendix B. In what 
follows we shall however show that the
threefold (\ref{TFOLD}) is birationally equivalent to a rational variety. As a consequence 
we will be  able to express both   
the transition operator and the $\mathrm{R}$-matrix without the need 
of any algebraic constraint.

\section{The Threefold Geometry}
\label{sectionGEOM}

The understanding of the geometrical properties of  
a specific algebraic variety require in general
its desingularization and this problem can be very difficult
for high dimensional manifolds \cite{REID}. Despite of the fact that
the threefold (\ref{TFOLD}) is not normal because of the presence of
two-dimensional singularities we have
been able to resolve them with the help of birational morphisms.
The concept of birational equivalence
is unique to Algebraic Geometry and through these mappings basic
invariants of the variety are preserved such as the generalization of the
concept of geometric genus \cite{HAR}.
We recall that the singular locus
of this threefold
is determined by the zeroes set of first order partial derivatives on the variables
$a_{+},b_{+},\bar{b}_{+},c_{+}$ and $c_{-}$. We find that they are constituted 
of three quadric surfaces,
\EQ
\label{SING}
\mathrm{Sing(T)}=\{[a_{+}:b_{+}:\bar{b}_{+}:c_{+}:c_{-}] \in \mathbb{CP}^4 | 
\phi_1(b_{+},\bar{b}_{+},c_{+})=0,\phi_2(a_{+},\bar{b}_{+})=0~\cup~a_{+}=0~\cup~c_{-}=0 \},
\EN
where the polynomials 
$\phi_1(b_{+},\bar{b}_{+},c_{+})=0$ and $\phi_2(a_{+},\bar{b}_{+})$
are,
\EQ
\label{POLaux}
\phi_1(b_{+},\bar{b}_{+},c_{+})= c_{+}^2 - b_{+}\bar{b}_{+},~~ 
\phi_2(a_{+},\bar{b}_{+})=\Lambda_2^{+}a_{+}+\Lambda_1^{+}\bar{b}_{+}.
\EN

The partial desingularization can be implemented by observing
that the threefold (\ref{TFOLD}) has two factorized terms both
containing quadratic forms. By exploring this fact we are able
to decrease the dimensionality of the singular locus and 
the degree of the image polynomial. This is done 
with the help of the following rational map,
\EQ
\label{mapT1}
\renewcommand{\arraystretch}{1.5}
\begin{array}{ccc}
\mathrm{T}(a_{+},b_{+},\bar{b}_{+},c_{+},c_{-}) \subset \mathbb{CP}^4 &~~~ \overset{\phi}{\longrightarrow}~~~ 
& \mathrm{T}_1(a_{+},b_{+},\bar{b}_{+},c_{+},c_{-}) \subset \mathbb{CP}^4 \\
\left[a_{+}:b_{+}:\bar{b}_{+}:c_{+}:c_{-}\right] & \longmapsto & \left[a_{+}:
b_{+}:\bar{b}_{+}
:c_{+}: \frac{a_{+}c_{-}\phi_2(a_{+},\bar{b}_{+})}{\phi_1(b_{+},\bar{b}_{+},c_{+})}\right],
\end{array}
\EN
where the map image is the cubic threefold,
\EQ
\mathrm{T}_1(a_{+},b_{+},\bar{b}_{+},c_{+},c_{-})=
\Lambda_2^{+}\left[c_{+}^2-b_{+}\bar{b}_{+}\right] \left[\Lambda_2^{+}a_{+}+\Lambda_1^{+}\bar{b}_{+}\right]
-a_{+}c_{-}^2-\Lambda_{1}^{-}\Lambda_2^{+}a_{+}^2b_{+}-\left([\Lambda_2^{+}]^2-1\right)a_{+}b_{+}\bar{b}_{+}.
\EN

We note that the map $\phi$ is everywhere defined except at 
the closed subsets consisted of 
the singular locus (\ref{SING}) which is allowed by the notion of birational
equivalence. 
Informally speaking, two varieties
are considered birational if they are isomorphic up to
lower dimensional subsets. This is exactly the case of 
the map $\phi$ since it is bijective 
over the non-singular 
locus of the threefold (\ref{TFOLD}) with the inverse,
\EQ
\label{mapT1INV}
\renewcommand{\arraystretch}{1.5}
\begin{array}{ccc}
\mathrm{T}_1(a_{+},b_{+},\bar{b}_{+},c_{+},c_{-}) \subset \mathbb{CP}^4 &~~~ \overset{\phi^{-1}}{\longrightarrow}~~~ 
& \mathrm{T}(a_{+},b_{+},\bar{b}_{+},c_{+},c_{-}) \subset \mathbb{CP}^4 \\
\left[a_{+}:b_{+}:\bar{b}_{+}:c_{+}:c_{-}\right] & \longmapsto & \left[a_{+}:b_{+}:\bar{b}_{+}:c_{+}:
\frac{c_{-}\phi_1(b_{+},\bar{b}_{+},c_{+})}{a_{+}\phi_2(a_{+},\bar{b}_{+})}\right].
\end{array}
\EN

Although the cubic threefold 
$\mathrm{T}_1$ is still 
singular we have drastically reduced the dimension of its singular locus. In fact, the singularities are now
made of three isolated points with the following 
projective coordinates, 
\EQ
\mathrm{P}_{1}=[0:1:0:0:0]~~\mathrm{and}~~\mathrm{P}_{\pm}=
[1:0:\frac{1-2[\Lambda_{2}^{+}]^2 \pm \mathrm{i} \sqrt{3}}{2\Lambda_1^{+} \Lambda_2^{+}}:0:0].
\EN

It is not difficult to show that any cubic threefold with isolated singularities 
is birationally equivalent to projective space $\mathbb{CP}^3$.
This fact is a consequence of the property that 
a generic set of lines passing through a given singular point will 
have a further intersection point with the cubic threefold.
Choosing $\mathrm{P}_1$ as the singular point 
we can build the rational map,
\EQ
\label{mapT2}
\renewcommand{\arraystretch}{1.5}
\begin{array}{ccc}
\mathbb{CP}^3 &~~~ \overset{\psi}{\longrightarrow}~~~ 
& \mathrm{T}_1(a_{+},b_{+},\bar{b}_{+},c_{+},c_{-}) \subset \mathbb{CP}^4 \\
\left[a_{+}:\bar{b}_{+}:c_{+}:c_{-}\right] & \longmapsto & \left[a_{+}:\frac{\psi_1(a_{+},\bar{b}_{+},c_{+},c_{-})}{\psi_2(a_{+},\bar{b}_{+})}:\bar{b}_{+}:c_{+}:
c_{-}\right],
\end{array}
\EN
where the expressions of the polynomials  $\psi_1(a_{+},\bar{b}_{+},c_{+},c_{-})$ and $\psi_2(a_{+},\bar{b}_{+})$ are,
\begin{eqnarray}
&&\psi_1(a_{+},\bar{b}_{+},c_{+},{c}_{-})=-a_{+}{c}_{-}^2+\Lambda_2^{+}\left[\Lambda_2^{+}a_{+}+\Lambda_1^{+}\bar{b}_{+}\right]c_{+}^2, \nonumber \\
&&\psi_2(a_{+},\bar{b}_{+})=\Lambda_1^{-}\Lambda_2^{+}a_{+}^2+\left(2[\Lambda_2^{+}]^2-1\right)a_{+}\bar{b}_{+}+\Lambda_1^{+}\Lambda_2^{+}\bar{b}_{+}^2.
\end{eqnarray}

The map $\psi$ is regular when restricted to dense open 
subsets of $\mathbb{CP}^3$ being once again birational.  
Its inverse is the projection,
\EQ
\label{mapT2INV}
\renewcommand{\arraystretch}{1.5}
\begin{array}{ccc}
\mathrm{T}_1(a_{+},b_{+},\bar{b}_{+},c_{+},c_{-}) \subset \mathbb{CP}^4 &~~~ \overset{\psi^{-1}}{\longrightarrow}~~~ 
& \mathbb{CP}^3 \\
\left[a_{+}:b_{+}:\bar{b}_{+}:c_{+}:c_{-}\right] & \longmapsto & \left[a_{+}:\bar{b}_{+}:c_{+}:
c_{-}\right].
\end{array}
\EN

The composition of birational mappings keeps the birationality property
and consequently we have been able to stablish 
the following equivalence,
\EQ
\mathrm{T}(a_{+},b_{+},\bar{b}_{+},c_{+},c_{-}) \backslash \ \mathrm{Sing(T)} 
\cong \mathbb{CP}^3.
\EN

We shall next use the above correspondence to 
exhibit a rational parameterization of the
transition operator. 

\subsection{Parameterized Transition Operator }
\label{subPARA}

We have now the basic ingredients to present an explicit expression 
for the transition operator
without any algebraic constraint among the weights. The map $\phi$ 
compels us to  introduce 
a new variable 
$\mathfrak{c}_{-}$ as follows, 
\EQ
\mathfrak{c}_{-}=\frac{a_{+}c_{-}
\phi_2(a_{+},\bar{b}_{+})}{
\phi_1(b_{+},\bar{b}_{+},c_{+})}, 
\EN
which now replaces $c_{-}$ as an independent variable.

The second map $\psi$ makes it possible the elimination of the weight
$b_{+}$. Taking into account
these steps together we are able to extract in a linear way the 
weights $b_{+}$ and $c_{-}$, 
\EQ
\label{auxwbc}
b_{+}=\frac{\psi_1(a_{+},\bar{b}_{+},c_{+},\mathfrak{c}_{-})}{\psi_2(a_{+},\bar{b}_{+})},~~
c_{-}=\mathfrak{c}_{-}\frac{\left[c_{+}^2\psi_2(a_{+},\bar{b}_{+})-\bar{b}_{+}\psi_1(a_{+},\bar{b}_{+},c_{+},
\mathfrak{c}_{-})\right]}{a_{+}\phi_2(a_{+},\bar{b}_{+})\psi_2(a_{+},\bar{b}_{+})}.
\EN

After these transformations the transition operator elements
become parameterized by four projective weights that is
$a_{+},\bar{b}_{+},c_{+}$ and $\mathfrak{c}_{-}$. Since one of them
can be used as normalization
this solution has three affine independent spectral parameters. 
Considering the matrix structure (\ref{LAX1}) we can represent
the transition operator as,
\EQ
\mathrm{L}(\omega)=\left[
\begin{array}{ccc|ccc|ccc}
 a_{+} & 0 & 0 & 0 & 0 & 0 & 0 & 0 & 0 \\
 0 & b_{+} & 0 & c_{+} & 0 & 0 & 0 & 0 & 0 \\
 0 & 0 & f_{+} & 0 & d_{+} & 0 & h & 0 & 0 \\ \hline
 0 & c_{+} & 0 & \bar{b}_{+} & 0 & 0 & 0 & 0 & 0 \\
 0 & 0 & d_{+} & 0 & g & 0 & d_{-} & 0 & 0 \\
 0 & 0 & 0 & 0 & 0 & \bar{b}_{-} & 0 & c_{-} & 0 \\\hline
 0 & 0 & h & 0 & d_{-} & 0 & f_{-} & 0 & 0 \\
 0 & 0 & 0 & 0 & 0 & c_{-} & 0 & b_{-} & 0 \\
 0 & 0 & 0 & 0 & 0 & 0 & 0 & 0 & a_{-} \\
\end{array}
\right],
\label{LAXG}
\EN
where the components of the set $\omega$ are
the free variables $a_{+},\bar{b}_{+},c_{+}$ and $\mathfrak{c}_{-}$. 

The other entries of the transition operator are fixed by the components of the set $\omega$  
with the help of Eqs.(\ref{BB1}-\ref{BB2}).
By substituting the weights $b_{+}$ and $c_{-}$ (\ref{auxwbc}) in the relations (\ref{BB1}-\ref{BB2})
and after performing few simplifications we obtain,
\begin{eqnarray}
&& d_{+}=\frac{c_{+}\psi_1(a_{+},\bar{b}_{+},c_{+},\mathfrak{c}_{-})}{\psi_2(a_{+},\bar{b}_{+})\phi_2(a_{+},\bar{b}_{+})},~~
d_{-}=\frac{\mathfrak{c}_{-}\bar{b}_{+}}{\Lambda_2^{+}\phi_2(a_{+},\bar{b}_{+})},~~
h=\frac{\mathfrak{c}_{-}c_{+}}{\phi_2(a_{+},\bar{b}_{+})}, \\ \nonumber \\
&&f_{+}=\frac{\psi_1(a_{+},\bar{b}_{+},c_{+},\mathfrak{c}_{-})\left[\phi_2(a_{+},\bar{b}_{+})\psi_1(a_{+},\bar{b}_{+},c_{+},\mathfrak{c}_{-})-\Lambda_1^{+}c_{+}^2\psi_2(a_{+},\bar{b}_{+})\right]}{a_{+}\phi_2(a_{+},\bar{b}_{+})[\psi_2(a_{+},\bar{b}_{+})]^2}, \\ \nonumber \\
&& b_{-}=\frac{\bar{b}_{+}\left[\Lambda_1^{-}\Lambda_2^{+}a_{+}c_{+}^2+\left[\mathfrak{c}_{-}^2+([\Lambda_2^{+}]^2-1)c_{+}^2\right]\bar{b}_{+}\right]}{\Lambda_2^{+}\phi_2(a_{+},\bar{b}_{+})\psi_2(a_{+},\bar{b}_{+})}, \\ \nonumber \\
&& \bar{b}_{-}=\frac{\psi_1(a_{+},\bar{b}_{+},c_{+},\mathfrak{c}_{-})\left[c_{+}^2\psi_2(a_{+},\bar{b}_{+})-\bar{b}_{+}\psi_1(a_{+},\bar{b}_{+},c_{+},\mathfrak{c}_{-})\right]}{\Lambda_2^{+}a_{+}\phi_2(a_{+},\bar{b}_{+})[\psi_2(a_{+},\bar{b}_{+})]^2}, \\ \nonumber \\
&& f_{-}=-\frac{\bar{b}_{+}\left[\Lambda_1^{-}\Lambda_2^{+}a_{+}+([\Lambda_2^{+}]^2-1)\bar{b}_{+}\right]}{\Lambda_2^{+}\phi_2(a_{+},\bar{b}_{+})}, \\ \nonumber \\
&& a_{-}=\frac{\left[c_{+}^2\psi_2(a_{+},\bar{b}_{+})-\bar{b}_{+}\psi_1(a_{+},\bar{b}_{+},c_{+},\mathfrak{c}_{-})\right] \left[\Lambda_1^{-}\Lambda_2^{+}a_{+}\mathfrak{c}_{-}^2+([\Lambda_2^{+}\mathfrak{c}_{-}]^2-c_{+}^2)\bar{b}_{+}\right]}{\Lambda_2^{+}a_{+}\phi_2(a_{+},\bar{b}_{+})
[\psi_2(a_{+},\bar{b}_{+})]^2}, \\ \nonumber \\
&& g=\frac{c_{+}^2}{a_{+}}+\frac{\bar{b}_{+}\psi_1(a_{+},\bar{b}_{+},c_{+},\mathfrak{c}_{-})
\left[a_{+}-\Lambda_2^{+}\phi_2(a_{+},\bar{b}_{+})\right]}{\Lambda_2^{+}a_{+}\phi_2(a_{+},\bar{b}_{+})\psi_2(a_{+},\bar{b}_{+})}.
\end{eqnarray}

The parameterized expressions for the corresponding $\mathrm{R}$-matrix elements are somehow
involved since contains six free affine spectral parameters. For sake of completeness they have been 
also explicitely exhibited in Appendix B. 

We conclude by noticing that the above transition operator is regular since
there exists a special value $\omega_0$ 
of the spectral variables such that  $\mathrm{L}(\omega_0)$ becomes 
the permutator. Direct inspection of the transition operator  entries
reveals us that the projective coordinates of this point are,
\EQ
\omega_0=[a_{+}:\bar{b}_{+}:c_{+}:\mathfrak{c}_{-}]=[1:0:1:\Lambda_{2}^{+}].
\EN
making it possible the construction of local conserved charges being
the Hamiltonian of the respective spin chain one of them. 

\subsection{The Spin-$1$ Hamiltonian }
\label{subHAMIL}

The Hamiltonian limit of the vertex model is obtained 
by expanding the logarithm of the transfer matrix 
around the 
regular point of the transition operator. Here we shall discuss 
the form of this operator in a general situation which is able to cover
potential interesting submanifolds. To this end it is wise to consider
the perturbation around the regular point directly on the weights of the
degree seven 
threefold. The expansion giving the permutator 
at zero order is as follows, 
\EQ
a_{+}=1 +\epsilon \dot{a}_{+},~~ 
b_{+}=0+\epsilon \dot{b}_{+},~~ 
\bar{b}_{+}=0+\epsilon \dot{\bar{b}}_{+},~~ 
c_{+}=1 +\epsilon \dot{c}_{+},~~ 
c_{-}=1 +\epsilon \dot{c}_{-},
\label{EXP}
\EN
where $\epsilon$ denotes the expansion parameter. 

The coefficients $\dot{a}_{+}, 
\dot{b}_{+}, 
\dot{\bar{b}}_{+}, 
\dot{c}_{+}$ 
and $\dot{c}_{-}$ need to satisfy  
the threefold polynomial (\ref{TFOLD}) up to 
the first order in the parameter $\epsilon$. This requires
that they are constrained by the relation,
\EQ
\Lambda_1^{-}\dot{b}_{+}+\Lambda_1^{+}\dot{\bar{b}}_{+}+4\Lambda_2^{+}\dot{a}_{+}
+2\Lambda_2^{+}\dot{c}_{-}-6\Lambda_2^{+}\dot{c}_{+}=0.
\label{CONHAM}
\EN

We shall see that combinations of these coefficients will play the role of two
additional coupling 
constants for the Hamiltonian. This is a manifestation of the fact that 
the vertex model we are considering sits on a three-dimensional variety. 
To obtain the spin chain expression
we have to expand the transition operator up to the first order on the
parameter $\epsilon$. The technical details 
of this computation have been summarized in
Appendix $\mathrm{C}$ and in what follows we present only the final result.
We find that the Hamiltonian is given by,
\begin{eqnarray}
\label{HAM}
\mathrm{H}&=&\sum_{j=1}^{\mathrm{N}}\Bigg\{(S_{j}^{-}S_{j+1}^{+}
+\frac{\dot{\bar{b}}_{+}}{\dot{b}_{+}}S_{j}^{+}S_{j+1}^{-})
(1+\frac{[\Lambda_2^{+}-1]^2}{\Lambda_2^{+}}S_{j}^{z}S_{j+1}^{z})
-\frac{\Lambda_1^{+}}{2}[S_{j}^{-}S_{j+1}^{+}]^2-\frac{\Lambda_1^{-}
\dot{\bar{b}}_{+}}{2\dot{b}_{+}}[S_{j}^{+}S_{j+1}^{-}]^2 \nonumber \\
&+&[\Lambda_2^{+}-1]\left(S_{j}^{-}S_{j}^{z}S_{j+1}^{+}+
\frac{\dot{\bar{b}}_{+}}{\dot{b}_{+}}S_{j}^{+}S_{j+1}^{-}S_{j+1}^{z}\right)
+[\frac{\Lambda_2^{+}-1}{\Lambda_2^{+}}]\left(S_{j}^{-}S_{j+1}^{+}S_{j+1}^{z}+
\frac{\dot{\bar{b}}_{+}}{\dot{b}_{+}}S_{j}^{+}S_{j}^{z}S_{j+1}^{-}\right)\nonumber \\
&-&(\Lambda_1^{-}+\frac{\Lambda_1^{+}\dot{\bar{b}}_{+}}{\dot{b}_{+}})[S_{j}^{z}]^2 
+2\frac{\Lambda_2^{+}}{\dot{b}_{+}}(\dot{c}_{+}-\dot{c}_{-})S_{j}^{z}
\Bigg\},
\end{eqnarray}
where $S_{j}^{+},S_{j}^{-}$ and $S_j^{z}$ are 
the spin-$1$ matrices of the 
$\mathrm{SU}(2)$ algebra.

In order to compare the Hamiltonian (\ref{HAM}) with the
one found previously by
Cramp\'e, Frappat and Ragoucy \cite{CFR}
we just have to apply a gauge transformation
on our solution of the Yang-Baxter equation.
This correspondence together with
the explicit matching of the Hamiltonian couplings
can also be found in
Appendix $\mathrm{C}$. 

An interesting feature of deriving the 
Hamiltonian from
the integrable vertex model is that it tells us 
the presence of an
arbitrary azimuthal magnetic field is directly connected
with the asymmetry  among the 
weights $c_{+}$ and $c_{-}$. This means that 
the simplest way to obtain a Hamiltonian in
absence of any linear magnetic field is to consider 
the threefold subvariety,
\EQ
c_{-}-c_{+}=0.
\label{sub}
\EN

The physical properties of the spin chain is expected to be dependent
whether or not we have the presence of the azimuthal magnetic field.
It is conceivable that this fact should be reflected as a drastic 
change of the geometrical properties
of the variety underlying the vertex model. We shall next confirm this
feeling by investigating the geometry of the 
hyperplane section (\ref{sub}). 

\section{The Integrable Submanifold }
\label{sectionSUBMANI}

The weights of the vertex model whose Hamiltonian limit does not contain
any azimuthal magnetic field sits on the intersection of $c_{+}-c_{-}=0$ with
the original threefold (\ref{TFOLD}). This leads us to a surface defined
by the polynomial,
\begin{eqnarray}
\label{SURF}
\mathrm{S}(a_{+},b_{+},\bar{b}_{+},c_{+})&=&
(c_{+}^2-b_{+}\bar{b}_{+})^2 \Bigg[ 
\Lambda_2^{+}(c_{+}^2-b_{+}\bar{b}_{+})(\Lambda_2^{+}a_{+}+\Lambda_1^{+}\bar{b}_{+})
-\Lambda_{1}^{-}\Lambda_2^{+}a_{+}^2b_{+} \nonumber \\ 
&-& ([\Lambda_2^{+}]^2-1)a_{+}b_{+}\bar{b}_{+} \Bigg]
-a_{+}^3c_{+}^2\left[\Lambda_2^{+}a_{+} + \Lambda_1^{+}\bar{b}_{+}\right]^2, 
\end{eqnarray}
which has two non coplanar singular lines given by,
\EQ
\mathrm{Sing(S)}=\{[a_{+}:0:\bar{b}_{+}:0] \cup [a_{+}:b_{+}:0:0] \}.
\EN

The classification of surfaces has been mostly done by
Enriques who divided them into four basic families according
to what nowadays is called Kodaira 
dimension. 
In
order to find out which family 
the above surface belongs we have to perform its
desingularization again with
the help of birational maps. As before the basic idea is lower
the dimensionality of the singular locus as well as the degree
of the image polynomial. Inspired by what has
been done for the threefold we are able to propose
the following rational map, 
\EQ
\label{mapS1}
\renewcommand{\arraystretch}{1.5}
\begin{array}{ccc}
\mathrm{S}(a_{+},b_{+},\bar{b}_{+},c_{+}) \subset \mathbb{CP}^3 &~~~ \overset{\sigma}{\longrightarrow}~~~ 
& \mathrm{S}_1(a_{+},b_{+},\bar{b}_{+},c_{+}) \subset \mathbb{CP}^3 \\
\left[a_{+}:b_{+}:\bar{b}_{+}:c_{+}\right] & \longmapsto & \left[a_{+}:
\frac{a_{+}c_{+}\phi_2(a_{+},\bar{b}_{+})}{\phi_1(b_{+},\bar{b}_{+},c_{+})}:\bar{b}_{+}
:c_{+}\right],
\end{array}
\EN
where the respective target variety is a surface of degree four,
\begin{eqnarray}
\label{SURF1}
\mathrm{S}_1(a_{+},b_{+},\bar{b}_{+},c_{+}) &=&
\Lambda_1^{-}\Lambda_2^{+}(\Lambda_2^{+}a_{+}^2-b_{+}c_{+})a_{+}c_{+}+\Lambda_2^{+}(\Lambda_1^{-}\Lambda_1^{+}+
2[\Lambda_2^{+}]^2-1)a_{+}^2{\bar{b}}_{+}c_{+} \nonumber \\
&+&(1-[\Lambda_2^{+}]^2)b_{+}{\bar{b}}_{+}c_{+}^2 
+\Lambda_1^{+}(3[\Lambda_2^{+}]^2-1)a_{+}{\bar{b}}_{+}^2c_{+}
+[\Lambda_1^{+}]^2\Lambda_2^{+}{\bar{b}}_{+}^3c_{+}-b_{+}^3{\bar{b}}_{+}. \nonumber \\
\end{eqnarray}

We see that the above map satisfies the expected properties of a birational equivalence onto its image.
In fact, the map $\sigma$ only fails on lower dimensional subsets
of the surface (\ref{SURF}) and it is easily invertible,   
\EQ
\label{mapS1INV}
\renewcommand{\arraystretch}{1.5}
\begin{array}{ccc}
\mathrm{S}_1(a_{+},b_{+},\bar{b}_{+},c_{+}) \subset \mathbb{CP}^3 &~~~ \overset{\sigma^{-1}}{\longrightarrow}~~~ 
& \mathrm{S}(a_{+},b_{+},\bar{b}_{+},c_{+}) \subset \mathbb{CP}^3 \\
\left[a_{+}:b_{+}:\bar{b}_{+}:c_{+}\right] & \longmapsto & \left[a_{+}:
\frac{b_{+}c_{+}^2-a_{+}c_{+}\phi_2(a_{+},\bar{b}_{+})}{b_{+}\bar{b}_{+}}:\bar{b}_{+}
:c_{+}\right].
\end{array}
\EN

We now are left to analyze the geometrical properties 
of the birationally equivalent
surface (\ref{SURF1}). This surface is fortunately normal because
its singularities consist of four isolated points 
with the following projective
coordinates,
\EQ
\label{SINGP}
\mathrm{P}_1=[0:0:0:1],~~\mathrm{P}_2=[1:0:-\frac{\Lambda_2^{+}}{\Lambda_1^{+}}:0],
~~\mathrm{and}~~\mathrm{P}_{\pm}=
[1:0:\frac{1-2[\Lambda_{2}^{+}]^2 \pm \mathrm{i} \sqrt{3}}{2\Lambda_1^{+} \Lambda_2^{+}}:0].
\EN

It is known that the minimal model
of a normal quartic surface can be either the celebrated $\mathrm{K}3$
surface, a ruled surface over an elliptic curve or still a rational
surface \cite{UME}. In order to decide the actual class 
of the surface (\ref{SURF1}) we have to investigate 
the local topological behavior of the
singular points according to 
the classification by Arnold et al. \cite{ANAR}. This problem can be sorted out
for example within the computer algebra system $\mathrm{Singular}$ \cite{SIN} which
enables us to compute the singularities modality. We found out that the singular
points (\ref{SINGP}) are canonical rational double points whose local forms are that of
the so-called $\mathrm{A}_2$ singularities \cite{ANAR}. As long as the
singularities are rational the minimal resolutions of singular quartics
are known to be $\mathrm{K}3$ 
surfaces see for instance \cite{URA}. Technically speaking, such resolutions do not affect adjunction and
the geometric properties are equivalent to that of a smooth quartic which is the simplest type of
$\mathrm{K}3$ surface. This discussion together with the map $\sigma$
are sufficient to characterize
the birational class of the degree seven surface (\ref{SURF}) that is,
\EQ
\mathrm{S}(a_{+},b_{+},\bar{b}_{+},c_{+}) \backslash \mathrm{Sing(S)}
\cong \mathrm{K}3~\mathrm{surface}.
\EN

For practical purposes it is of clear interest to have
the explicit form of the transition operator 
with weights constrained 
by the $\mathrm{K}3$ surface with the 
lowest possible degree. This can be achieved with the help of 
the birational map $\sigma$ which makes it possible  
to replace the weight $b_{+}$ by a new variable 
$\mathfrak{b}_{+}$ as follows,
\EQ
\label{auxwbb}
b_{+}=\frac{c_{+}\left[\mathfrak{b}_{+}c_{+}-a_{+}\phi_2(a_{+},\bar{b}_{+})\right]}{\mathfrak{b}_{+}\bar{b}_{+}}.
\EN

After this transformation the transition operator elements will become dependent on the projective
coordinates
$a_{+},\mathfrak{b}_{+},\bar{b}_{+}$ and $c_{+}$. Its matrix representation is as before except that
now $c_{-}=c_{+}$, namely
\EQ
\mathrm{L}(\omega)=\left[
\begin{array}{ccc|ccc|ccc}
 a_{+} & 0 & 0 & 0 & 0 & 0 & 0 & 0 & 0 \\
 0 & b_{+} & 0 & c_{+} & 0 & 0 & 0 & 0 & 0 \\
 0 & 0 & f_{+} & 0 & d_{+} & 0 & h & 0 & 0 \\ \hline
 0 & c_{+} & 0 & \bar{b}_{+} & 0 & 0 & 0 & 0 & 0 \\
 0 & 0 & d_{+} & 0 & g & 0 & d_{-} & 0 & 0 \\
 0 & 0 & 0 & 0 & 0 & \bar{b}_{-} & 0 & c_{+} & 0 \\\hline
 0 & 0 & h & 0 & d_{-} & 0 & f_{-} & 0 & 0 \\
 0 & 0 & 0 & 0 & 0 & c_{+} & 0 & b_{-} & 0 \\
 0 & 0 & 0 & 0 & 0 & 0 & 0 & 0 & a_{-} \\
\end{array}
\right],
\label{LAXG1}
\EN
where now $\omega$ stands for the variables 
$a_{+},\mathfrak{b}_{+},\bar{b}_{+}$ and $c_{+}$.

The respective matrix elements can be computed by
substituting the transformation (\ref{auxwbb}) in the general 
relations (\ref{BB1}-\ref{BB2}) and as a result we obtain,
\begin{eqnarray}
&&d_{+}=c_{+}^2\frac{\left[\mathfrak{b}_{+}c_{+}-a_{+}\phi_2(a_{+},\bar{b}_{+})\right]}{\mathfrak{b}_{+}\bar{b}_{+}\phi_2(a_{+},\bar{b}_{+})},~~
b_{-}=\frac{\bar{b}_{+}c_{+}}{\Lambda_2^{+}\mathfrak{b}_{+}},~~ 
d_{-}=\frac{\mathfrak{b}_{+}\bar{b}_{+}}{\Lambda_2^{+}\phi_2(a_{+},\bar{b}_{+})}, \\ \nonumber \\
&&f_{+}=\frac{c_{+}^2\left[\mathfrak{b}_{+}c_{+}-a_{+}\phi_2(a_{+},\bar{b}_{+})\right]\left[\Lambda_2^{+}
\mathfrak{b}_{+}c_{+}-\phi_2(a_{+},\bar{b}_{+})^2\right]}{[\mathfrak{b}_{+}\bar{b}_{+}]^2\phi_2(a_{+},\bar{b}_{+})}, \\ \nonumber \\
&&g=\frac{c_{+}[\mathfrak{b}_{+}c_{+}-a_{+}\phi_2(a_{+},\bar{b}_{+})+\Lambda_2^{+}\phi_2(a_{+},
\bar{b}_{+})^2]}{\Lambda_2^{+}\mathfrak{b}_{+}\phi_2(a_{+},\bar{b}_{+})},~~
h=\frac{\mathfrak{b}_{+}c_{+}}{\phi_2(a_{+},\bar{b}_{+})}, \\ \nonumber \\
&&a_{-}=\frac{c_{+}^2\left([\Lambda_2^{+}\bar{b}_{+}+\Lambda_1^{-}a_{+}]\phi_2(a_{+},\bar{b}_{+})
-\Lambda_1^{-}\mathfrak{b}_{+}c_{+}\right)}{\mathfrak{b}_{+}^2\bar{b}_{+}},~~
\bar{b}_{-}=\frac{c_{+}^2\left[\mathfrak{b}_{+}c_{+}-a_{+}\phi_2(a_{+},\bar{b}_{+})\right]}{\Lambda_2^{+}\mathfrak{b}_{+}^2\bar{b}_{+}}, \\ \nonumber \\
&&f_{-}=\frac{\bar{b}_{+}^2\left[\bar{b}_{+}c_{+}\phi_2(a_{+},\bar{b}_{+})-\Lambda_1^{-}\Lambda_2^{+}\mathfrak{b}_{+}^3\right]}{[\Lambda_2^{+}]^2c_{+}\phi_2(a_{+},\bar{b}_{+})\left([\Lambda_2^{+}\bar{b}_{+}+\Lambda_1^{-}a_{+}]\phi_2(a_{+},\bar{b}_{+})-\Lambda_1^{-}\mathfrak{b}_{+}c_{+}\right)}.
\end{eqnarray}

Here we emphasize that the variables 
$a_{+},\mathfrak{b}_{+},\bar{b}_{+}$ and $c_{+}$ are
constrained by the quartic $\mathrm{K}3$ surface 
$\mathrm{S}_1(a_{+},\mathfrak{b}_{+},\bar{b}_{+},c_{+})$. In other 
words the polynomial 
given by Eq.(\ref{SURF1}) in which the
variable $b_{+}$ is replaced by the new auxiliary 
weight $\mathfrak{b}_{+}$. We finally remark that  
the respective $\mathrm{R}$-matrix entries as function of such 
variables have been presented in Appendix B. Although we have
the symmetry $c_{+}=c_{-}$ for the transition operator the same
does not occur for the equivalent entries of the $\mathrm{R}$-matrix.
This probably is related to the fact that $\mathrm{K}3$ surfaces do
not have an underlying group structure.

\section{Conclusions}
\label{sectionCON}

In this work we have discussed some guidelines to search for 
solutions of the Yang-Baxter equation 
from the point of view of Algebraic Geometry. We have applied this approach 
in the case of nineteen vertex models with the time-reversal symmetry which restrict 
the parameters space to fourteen distinct Boltzmann weights. Even in this
subspace we have to deal with a large
number of independent functional relations being 
therefore a satisfactory test of the practical
utility of our framework. We have been able to uncover a family of 
such integrable models lying
on a degree seven algebraic threefold whose polynomial coefficients depends on two free 
couplings. By means of birational mappings we have shown that this variety is equivalent
to the projective space $\mathbb{CP}^3$. These transformations are used to express 
both the transition operator and the $\mathrm{R}$-matrix in a parameterized form with
three independent affine spectral variables. We have verified explicitly 
that the $\mathrm{R}$-matrix satisfies the Yang-Baxter equation for three
distinct sets of non-additive spectral parameters. 

We have discussed the Hamiltonian limit of this family of nineteen vertex
model and show how it can be related to the spin-$1$ chain discovered 
in the previous work \cite{CFR}. We have argued that the presence of
an arbitrary azimuthal magnetic field in 
the Hamiltonian comes from the fact that it is proportional to the 
asymmetry between two kinds of spectral weights. This prompted us to
define the physical interesting submanifold in which the magnetic
field is absent. It turns out that the geometric properties 
of such submanifold are on the class of the famous $\mathrm{K}3$ 
surfaces. This finding has enlarged in substantial way the type
of non-rational varieties that can emerge as solution of the Yang-Baxter
equation.
A natural question to be
asked is whether this is an isolated result or actually a tip of an iceberg?
At least from the perspective of Algebraic Geometry $\mathrm{K}3$ surfaces
can exist in vast quantities as 
certain sections of a family of threefolds called Fano \cite{REID1}. 
This suggests that the answer of the above question is somehow related
to the existence or not of an abundance of integrable models sitting on
three-dimensional Fano varieties. 

Finally, we believe that the algebraic approach discussed in this work could be
extended to provide the complete classification of three-states
vertex models invariant by the $\mathrm{U}(1)$ symmetry together
with the proper identification of the respective 
algebraic varieties. We hope 
to report on this problem in a forthcoming  work.

\section*{Acknowledgments}
This work has been support by the Brazilian Research Councils CNPq and FAPESP.

\addcontentsline{toc}{section}{Appendix A}
\section*{\bf Appendix A: Extra Functional Relations}
\setcounter{equation}{0}
\renewcommand{\theequation}{A.\arabic{equation}}

In the main text we have solved thirty-nine functional equations 
out of the fifty-seven relations coming
from the Yang-Baxter equation (\ref{YBA}). The remaining eighteen 
functional equations mixed several weights
and $\mathrm{R}$-matrix elements previously 
determined. Their explicit expression are given by,
\begin{eqnarray} 
\label{extraeqi}
&&{\bf{c}}_{\pm} d_{\mp}^{'} a_{\pm}^{''} - {\bf{h}} a_{\pm}^{'} d_{\mp}^{''} - {\bf{f}}_{\mp} h^{'} d_{\pm}^{''} - {\bf{d}}_{\mp} c_{\pm}^{'} g^{''}=0, \\
&&{\bf{a}}_{\pm} d_{\pm}^{'} c_{\pm}^{''} - {\bf{g}} c_{\pm}^{'} d_{\pm}^{''} - {\bf{d}}_{\mp} h^{'} f_{\pm}^{''} - {\bf{d}}_{\pm} a_{\pm}^{'} h^{''}=0, \\
&&{\bf{c}}_{\mp} f_{\mp}^{'} c_{\pm}^{''} - {\bf{d}}_{\mp} g^{'} d_{\mp}^{''} - {\bf{h}} c_{\pm}^{'} f_{\mp}^{''} - {\bf{f}}_{\mp} c_{\mp}^{'} h^{''}=0, \\
&&{\bf{c}}_{\pm} g^{'} c_{\pm}^{''} - {\bf{d}}_{\pm} a_{\pm}^{'} d_{\mp}^{''} - {\bf{d}}_{\mp} h^{'} d_{\pm}^{''} - {\bf{g}} c_{\pm}^{'} g^{''}  +
     {\bf{b}}_{\pm} c_{\pm}^{'} {\bar{b}}_{\pm}^{''} =0, \\
&&{\bf{h}} d_{\mp}^{'} c_{\pm}^{''} - {\bf{g}} c_{\pm}^{'} d_{\mp}^{''} - {\bf{d}}_{\pm} a_{\pm}^{'} f_{\mp}^{''} - {\bf{d}}_{\mp} h^{'} h^{''}  +
     {\bf{d}}_{\pm} {\bar{b}}_{\pm}^{'} {\bar{b}}_{\pm}^{''} =0, \\
&&{\bf{f}}_{\pm} a_{\pm}^{'} d_{\mp}^{''} - {\bf{b}}_{\pm} b_{\pm}^{'} d_{\mp}^{''} + {\bf{h}} h^{'} d_{\pm}^{''} + {\bf{d}}_{\pm} c_{\pm}^{'} g^{''} - 
     {\bf{c}}_{\pm} d_{\pm}^{'} h^{''}=0, \\
&&\overline{\bf{b}}_{\mp} c_{\pm}^{'} b_{\mp}^{''} + {\bf{c}}_{\mp} h^{'} c_{\mp}^{''} - {\bf{d}}_{\pm} g^{'} d_{\mp}^{''} - {\bf{f}}_{\pm} c_{\pm}^{'} f_{\mp}^{''} - 
     {\bf{h}} c_{\mp}^{'} h^{''}=0, \\
&&{\bf{g}} g^{'} d_{\mp}^{''} - \overline{\bf{b}}_{\mp} {\bar{b}}_{\pm}^{'} d_{\mp}^{''} + {\bf{d}}_{\pm} c_{\pm}^{'} f_{\mp}^{''}  -
     {\bf{c}}_{\mp} d_{\mp}^{'} g^{''} + {\bf{d}}_{\mp} c_{\mp}^{'} h^{''}=0, \\ 
\label{extraeqf}
&&{\bf{d}}_{\mp} b_{\mp}^{'} b_{\pm}^{''} + {\bf{g}} d_{\mp}^{'} c_{\pm}^{''} - {\bf{h}} c_{\pm}^{'} d_{\mp}^{''}  -
     {\bf{f}}_{\mp} c_{\mp}^{'} d_{\pm}^{''} - {\bf{d}}_{\mp} g^{'} g^{''}=0.
\end{eqnarray}

By substituting the data of the solution described in section (\ref{sectionYBE}) we are going
to end with eighteen polynomials depending solely on the weights $a_{+},b_{+},\bar{b}_{+},c_{+}$ and
$c_{-}$. The main point is that the dependence of such polynomials on the variable $c_{-}$ appears
only by means of even powers. It turns out that the lowest possible power $c_{-}^2$ can be extracted
from the threefold (\ref{TFOLD}).
Denoting this power by $\mathrm{aux}$ it is given by,
\begin{eqnarray}
\label{auxcm}
\mathrm{aux} &=&
\left[
(-1 + [\Lambda_2^{+}]^2 - [\Lambda_2^{+}]^4) a_{+}^2 b_{+}+ 
\Lambda_1^{+} [\Lambda_2^{+}]^2 (c_{+}^2 - b_{+} {\bar{b}}_{+}) (\Lambda_2^{+} a_{+}+\Lambda_1^{+} {\bar{b}}_{+}) 
+\Lambda_1^{+} \Lambda_2^{+} (1 - [\Lambda_2^{+}]^2) a_{+} b_{+} {\bar{b}}_{+}
\right] \nonumber \\
&\times&
\left[ \frac{(c_{+}^2 - b_{+} {\bar{b}}_{+})^2} 
{\Lambda_1^{+} \Lambda_2^{+} a_{+}^3 (\Lambda_2^{+} a_{+} + \Lambda_1^{+} {\bar{b}}_{+})^2}\right].
\end{eqnarray}

The main idea is to replace the even powers on the weight $c_{-}$ 
by the above auxiliary variable in a nested way 
starting from the highest power.
Direct inspection of the polynomials associated to 
Eqs.(\ref{extraeqi}-\ref{extraeqf}) reveals us that 
the highest power is $c_{-}^6$. Denoting a given functional equation 
by $\mathrm{eq}[*]$ we can replace the powers on the weight $c_{-}$ by using the 
following Mathematica code,
\begin{eqnarray}
\mathrm{eq}1 &=& \mathrm{Factor}[\mathrm{eq}[*]], \\
\mathrm{eq}2 &=& \mathrm{Factor}[\mathrm{eq}1 ~/.~ \{ [c^{'}]^6 \rightarrow ~\mathrm{aux}^{'}~ [c^{'}]^4, [c^{''}]^6 \rightarrow \mathrm{aux}^{''}~ [c^{''}]^4 \}], \\
\mathrm{eq}3 &=& \mathrm{Factor}[\mathrm{eq}2 ~/.~ \{ [c^{'}]^4 \rightarrow \mathrm{aux}^{'}~ [c^{'}]^2, [c^{''}]^4 \rightarrow \mathrm{aux}^{''}~ [c^{''}]^2 \}], \\
\mathrm{eqend} &=& \mathrm{Factor}[\mathrm{eq}3 ~/.~ \{ [c^{'}]^2 \rightarrow \mathrm{aux}^{'}, [c^{''}]^2 \rightarrow \mathrm{aux}^{''} \}], 
\end{eqnarray}
where $\mathrm{aux}^{'}$ and $\mathrm{aux}^{''}$ are given by Eq.(\ref{auxcm}) with weights
labeled by $'$ and $''$ respectively.  

It is not difficult to check that the simplified relation $\mathrm{eqend}$ is always zero. In this way we
are able to verify that the whole Yang-Baxter equation is algebraically verified.

\addcontentsline{toc}{section}{Appendix B}
\section*{\bf Appendix B: The $\mathrm{R}$-matrix }
\label{APENB}
\setcounter{equation}{0}
\renewcommand{\theequation}{B.\arabic{equation}}

The $\mathrm{R}$-matrix depends on both sets of
spectral variables $\omega$ and $\omega^{'}$ and without loss of generality it can be
normalized by the element $\bf{c}_{+}$. Its matrix representation becomes,
\EQ
\scriptsize{
\mathrm{R}(\omega,\omega^{'})=
\left[
\begin{array}{ccccccccc}
 \bf{a}_{+}(\omega,\omega^{'}) & 0 & 0 & 0 & 0 & 0 & 0 & 0 & 0 \\
 0 & \bf{b}_{+}(\omega,\omega^{'}) & 0 & 1 & 0 & 0 & 0 & 0 & 0 \\
 0 & 0 & \bf{f}_{+}(\omega,\omega^{'}) & 0 & \bf{d}_{+}(\omega,\omega^{'}) & 0 & \bf{h}(\omega,\omega^{'}) & 0 & 0 \\ 
 0 & 1 & 0 & \overline{\bf{b}}_{+}(\omega,\omega^{'}) & 0 & 0 & 0 & 0 & 0 \\
 0 & 0 & {\bf{d}}_{+}(\omega,\omega^{'}) & 0 & \bf{g}(\omega,\omega^{'}) & 0 & {\bf{d}}_{-}(\omega,\omega^{'}) & 0 & 0 \\
 0 & 0 & 0 & 0 & 0 & \overline{\bf{b}}_{-}(\omega,\omega^{'}) & 0 & {\bf{c}}_{-}(\omega,\omega^{'}) & 0 \\
 0 & 0 & \bf{h}(\omega,\omega^{'}) & 0 & \bf{d}_{-}(\omega,\omega^{'}) & 0 & \bf{f}_{-}(\omega,\omega^{'}) & 0 & 0 \\
 0 & 0 & 0 & 0 & 0 & \bf{c}_{-}(\omega,\omega^{'}) & 0 & \bf{b}_{-}(\omega,\omega^{'}) & 0 \\
 0 & 0 & 0 & 0 & 0 & 0 & 0 & 0 & \bf{a}_{-}(\omega,\omega^{'}) \\
\end{array}
\right].
}
\label{RMA1}
\EN

Let us first discuss the form of the $\mathrm{R}$-matrix 
in the generic situation 
where the components of the set $\omega$ are the 
variables $a_{+},b_{+},\bar{b}_{+},c_{+},c_{-}$ which satisfy 
the threefold polynomial (\ref{TFOLD}). 
We first eliminate the even powers on the variables 
$c_{-}$ and $c_{-}^{'}$ of the $\mathrm{R}$-matrix elements
as explained in the previous Appendix. After this step
we still end up with some complicated 
polynomials but we find that they can be expressed in closed forms
once we define the auxiliary bi-homogeneous polynomial,
\begin{eqnarray}
\mathrm{Q}(a_{+},a_{+}^{'},b_{+},b_{+}^{'},\bar{b}_{+},\bar{b}_{+}^{'},c_{+},c_{+}^{'})&=&
\Lambda_1^{-}\Lambda_2^{+}a_{+}b_{+}{a}_{+}^{'}{b}_{+}^{'}
+(1-[\Lambda_2^{+}]^2)\left(c_{+}^2{a}_{+}^{'}{b}_{+}^{'}-b_{+}\bar{b}_{+}{a}_{+}^{'}{b}_{+}^{'}\right) \nonumber \\
&-&[\Lambda_2^{+}]^2\left(a_{+}b_{+}[{c}_{+}^{'}]^2-a_{+}b_{+}{b}_{+}^{'}\bar{b}_{+}^{'}\right) \nonumber \\
&+&\Lambda_1^{+}\Lambda_2^{+}\left(c_{+}^2-b_{+}\bar{b}_{+}\right)\left([{c}_{+}^{'}]^2-{b}_{+}^{'}\bar{b}_{+}^{'}\right).
\end{eqnarray}

Considering the above polynomial the expressions of the $\mathrm{R}$-matrix are given by,
\begin{eqnarray}
&&{\bf{a}}_{+}(\omega,\omega^{'}) = \frac{\bar{b}_{+}{a}_{+}^{'}{b}_{+}^{'} + a_{+}\phi_1({b}_{+}^{'},\bar{b}_{+}^{'},c_{+}^{'})}{
     c_{+}{a}_{+}^{'}{c}_{+}^{'}},~~
\overline{\bf{b}}_{+}(\omega,\omega^{'}) = \frac{\bar{b}_{+}{a}_{+}^{'} - a_{+}\bar{b}_{+}^{'}}{c_{+}{c}_{+}^{'}} ,\\ \nonumber \\
&&{\bf{b}}_{+}(\omega,\omega^{'}) = \frac{a_{+}b_{+}\phi_1(b_{+}^{'},\bar{b}_{+}^{'},c_{+}^{'})
-\phi_1(b_{+},\bar{b}_{+},c_{+}){a}_{+}^{'}{b}_{+}^{'}} 
{a_{+}c_{+}{a}_{+}^{'}{c}_{+}^{'}}, \\ \nonumber \\
&& {\bf{d}}_{+}(\omega,\omega^{'})=-\frac{{\bf{b}}_{+}(\omega,\omega^{'})c_{+}[{a}_{+}^{'}]^2{c}_{-}^{'}\phi_2({a}_{+}^{'},\bar{b}_{+}^{'})}{\phi_1({b}_{+}^{'},\bar{b}_{+}^{'},c_{+}^{'})\mathrm{Q}(a_{+},{a}_{+}^{'},1,{b}_{+}^{'},\bar{b}_{+},\bar{b}_{+}^{'},0,{c}_{+}^{'})}, \\ \nonumber \\
&& {\bf{f}}_{+}(\omega,\omega^{'})=\frac{{\bf{b}}_{+}(\omega,\omega^{'})\mathrm{Q}(a_{+},{a}_{+}^{'},b_{+},{b}_{+}^{'},\bar{b}_{+},
\bar{b}_{+}^{'},c_{+},{c}_{+}^{'})}{a_{+}\mathrm{Q}(a_{+},{a}_{+}^{'},1,{b}_{+}^{'},\bar{b}_{+},\bar{b}_{+}^{'},0,{c}_{+}^{'})}, \\ \nonumber \\
&&{\bf{h}}(\omega,\omega^{'})=\frac{[{a}_{+}^{'}]^2{c}_{-}^{'}\phi_1(b_{+},\bar{b}_{+},c_{+})\phi_2({a}_{+}^{'},\bar{b}_{+}^{'})
\left[\mathrm{Q}(a_{+},a_{+},b_{+},1,\bar{b}_{+},\bar{b}_{+},c_{+},0)-a_{+}c_{+}^2\right]}
{a_{+}^2c_{-}\phi_1({b}_{+}^{'},\bar{b}_{+}^{'},c_{+}^{'})\phi_2(a_{+},\bar{b}_{+})
\mathrm{Q}(a_{+},{a}_{+}^{'},1,{b}_{+}^{'},\bar{b}_{+},\bar{b}_{+}^{'},0,{c}_{+}^{'})}, \\ \nonumber \\
&& {\bf{b}}_{-}(\omega,\omega^{'})=-\frac{\overline{\bf{b}}_{+}(\omega,\omega^{'}){a}_{+}^{'}
\left[a_{+}b_{+}\bar{b}_{+}^{'}+\phi_1(b_{+},\bar{b}_{+},c_{+})a_{+}^{'}\right]} 
{a_{+}\mathrm{Q}(a_{+},{a}_{+}^{'},1,{b}_{+}^{'},\bar{b}_{+},\bar{b}_{+}^{'},0,{c}_{+}^{'})},~~ 
\overline{\bf{b}}_{-}(\omega,\omega^{'})=\frac{{\bf{b}}_{+}(\omega,\omega^{'}){\bf{b}}_{-}(\omega,\omega^{'})}
{\overline{\bf{b}}_{+}(\omega,\omega^{'})}, \\ \nonumber \\
&& {\bf{c}}_{-}(\omega,\omega^{'})=-\frac{c_{-}[{a}_{+}^{'}]^2{c}_{-}^{'}\phi_2(a_{+},\bar{b}_{+})\phi_2({a}_{+}^{'},\bar{b}_{+}^{'})
\left[a_{+}b_{+}\bar{b}_{+}^{'}+\phi_1(b_{+},\bar{b}_{+},c_{+})a_{+}^{'}\right]} 
{c_{+}{c}_{+}^{'}\phi_1(b_{+},\bar{b}_{+},c_{+})\phi_1({b}_{+}^{'},\bar{b}_{+}^{'},c_{+}^{'})\mathrm{Q}(a_{+},{a}_{+}^{'},1,{b}_{+}^{'},\bar{b}_{+},\bar{b}_{+}^{'},0,{c}_{+}^{'})}, \\ \nonumber \\
&&{\bf{d}}_{-}(\omega,\omega^{'})=-\frac{\overline{\bf{b}}_{+}(\omega,\omega^{'})
a_{+}c_{-}{a}_{+}^{'}{c}_{+}^{'}
\phi_2(a_{+},\bar{b}_{+})}{\phi_1(b_{+},\bar{b}_{+},c_{+})
\mathrm{Q}(a_{+},{a}_{+}^{'},1,{b}_{+}^{'},\bar{b}_{+},\bar{b}_{+}^{'},0,{c}_{+}^{'})}, \\ \nonumber \\
&&{\bf{f}}_{-}(\omega,\omega^{'})=\frac{\overline{\bf{b}}_{+}(\omega,\omega^{'}){a}_{+}^{'}
\mathrm{Q}(a_{+},{a}_{+}^{'},1,1,\bar{b}_{+},\bar{b}_{+}^{'},0,0)}
{\mathrm{Q}(a_{+},{a}_{+}^{'},1,{b}_{+}^{'},\bar{b}_{+},\bar{b}_{+}^{'},0,{c}_{+}^{'})}, \\ \nonumber \\
&&{\bf{a}}_{-}(\omega,\omega^{'})=\frac{{a}_{+}^{'}\left[\mathrm{Q}(a_{+},{a}_{+}^{'},b_{+},1,\bar{b}_{+},\bar{b}_{+}^{'},c_{+},0)-
a_{+}b_{+}\bar{b}_{+}^{'}-\phi_1(b_{+},\bar{b}_{+},c_{+})a_{+}^{'}\right]}{ 
a_{+}^2c_{+}{c}_{+}^{'}\mathrm{Q}(a_{+},{a}_{+}^{'},1,{b}_{+}^{'},\bar{b}_{+},\bar{b}_{+}^{'},0,{c}_{+}^{'})} \nonumber \\
&&~~~~~~~~~~~~\times
\left[a_{+}b_{+}\bar{b}_{+}^{'}+\phi_1(b_{+},\bar{b}_{+},c_{+})a_{+}^{'}\right], \\ \nonumber \\ 
&&{\bf{g}}(\omega,\omega^{'})=-\frac{\overline{\bf{b}}_{+}(\omega,\omega^{'})\mathrm{Q}({a}_{+}^{'},a_{+},{b}_{+}^{'},b_{+},
\bar{b}_{+}^{'},\bar{b}_{+},{c}_{+}^{'},c_{+})}{a_{+}\mathrm{Q}(a_{+},{a}_{+}^{'},1,{b}_{+}^{'},\bar{b}_{+},\bar{b}_{+}^{'},0,{c}_{+}^{'})} 
+\frac{c_{+}\left[\mathrm{Q}({a}_{+}^{'},{a}_{+}^{'},{b}_{+}^{'},1,\bar{b}_{+}^{'},\bar{b}_{+}^{'},{c}_{+}^{'},0)-{a}_{+}^{'}[{c}_{+}^{'}]^2\right]}{{c}_{+}^{'}\mathrm{Q}(a_{+},{a}_{+}^{'},1,{b}_{+}^{'},\bar{b}_{+},\bar{b}_{+}^{'},0,{c}_{+}^{'})}, \nonumber \\
\end{eqnarray}
where the polynomials 
$\phi_1(b_{+},\bar{b}_{+},c_{+})$ and $\phi_2(a_{+},\bar{b}_{+})$ have been defined in Eq.(\ref{POLaux}).

It is not difficult to check that this 
$\mathrm{R}$-matrix indeed satisfy the regularity condition (\ref{INI}). By a systematic use of the threefold (\ref{TFOLD}) 
as explained in the previous Appendix we are able to verified 
it also satisfies the 
Yang-Baxter equation for non-additive
operators, namely
\EQ
\mathrm{R}_{12}(\omega,\omega^{'}) 
\mathrm{R}_{13}(\omega,\omega^{''}) 
\mathrm{R}_{23}(\omega^{'},\omega^{''})= 
\mathrm{R}_{23}(\omega^{'},\omega^{''}) 
\mathrm{R}_{13}(\omega,\omega^{''}) 
\mathrm{R}_{12}(\omega,\omega^{'}) 
\EN
defined on three distinct sets of weights lying on the threefold (\ref{TFOLD}).
As usual the subscript labels of the $\mathrm{R}$-matrix 
indicates its non-trivial action on the product of the three 
distinct auxiliary spaces. Recall that this relation is a sufficient 
condition for the
associativity of the quadratic
algebra we have started with (\ref{YBA}).

$\bullet$ {\bf Parameterized $\mathrm{R}$-matrix}

The threefold (\ref{TFOLD}) was shown to be birationally equivalent to
the projective space $\mathbb{CP}^3$ and therefore it is possible to 
present the $\mathrm{R}$-matrix entries as function of four free
parameters.
In this situation we recall that the set $\omega$ is constituted
of the independent variables $a_{+},\bar{b}_{+},c_{+},\mathfrak{c}_{-}$. 
Once again we find that the $\mathrm{R}$-matrix elements can be
better expressed with the help of an auxiliary polynomial, namely
\begin{eqnarray}
\mathrm{Q}_1({a}_{+},{a}_{+}^{'},\bar{b}_{+},{\bar{b}}_{+}^{'},x_1,x_2,y_1,y_2)&=&  
\left[\Lambda_1^{-}\Lambda_2^{+}a_{+}a_{+}^{'} + (-1 + [\Lambda_2^{+}]^2){\bar{b}}_{+}a_{+}^{'} + 
     \Lambda_2^{+}(\Lambda_2^{+}a_{+} + \Lambda_1^{+}{\bar{b}}_{+}){\bar{b}}_{+}^{'}\right](x_1y_2)^2 \nonumber \\
&+&\left[-\Lambda_1^{-}\Lambda_2^{+}a_{+}a_{+}^{'} + a_{+}{\bar{b}}_{+}^{'} - 
     \Lambda_2^{+}(\Lambda_2^{+}{\bar{b}}_{+}a_{+}^{'} + \Lambda_2^{+}a_{+}{\bar{b}}_{+}^{'} + \Lambda_1^{+}{\bar{b}}_{+}{\bar{b}}_{+}^{'})\right](x_2y_1)^2 \nonumber \\
&+&\left[{\bar{b}}_{+}a_{+}^{'} - a_{+}{\bar{b}}_{+}^{'}\right]\left[(x_1x_2)^2+(y_1y_2)^2\right],
\end{eqnarray}
where $x_1,x_2,y_1,y_2$ are parameters taken 
values on the specific set $0,1,\mathfrak{c}_{-},\mathfrak{c}_{-}^{'}$. 

Considering the above polynomial and after some cumbersome simplifications
we find that the parameterized expressions for the matrix elements are,
\begin{eqnarray}
&&{\bf{a}}_{+}(\omega,\omega^{'})=-\frac{\mathrm{Q}_1({a}_{+},{a}_{+}^{'},\bar{b}_{+},{\bar{b}}_{+}^{'},0,{c}_{+}^{'},1,{\mathfrak{c}}_{-}^{'})}{{c}_{+}{c}_{+}^{'}\psi_2({a}_{+}^{'},{\bar{b}}_{+}^{'})} 
,~~
{\bf{b}}_{+}(\omega,\omega^{'})=\frac{\mathrm{Q}_1({a}_{+},{a}_{+}^{'},\bar{b}_{+},{\bar{b}}_{+}^{'},{c}_{+},{c}_{+}^{'},{\mathfrak{c}}_{-},{\mathfrak{c}}_{-}^{'})}{{c}_{+}{c}_{+}^{'}\psi_2({a}_{+},\bar{b}_{+})\psi_2({a}_{+}^{'},{\bar{b}}_{+}^{'})} 
,\\ \nonumber \\
&&\overline{\bf{b}}_{+}(\omega,\omega^{'})=\frac{\bar{b}_{+}{a}_{+}^{'} - {a}_{+}{\bar{b}}_{+}^{'}}{{c}_{+}{c}_{+}^{'}} 
,~~
{\bf{d}}_{+}(\omega,\omega^{'})=\frac{{\mathfrak{c}}_{-}^{'}\mathrm{Q}_1({a}_{+},{a}_{+}^{'},\bar{b}_{+},{\bar{b}}_{+}^{'},{c}_{+},{c}_{+}^{'},{\mathfrak{c}}_{-},{\mathfrak{c}}_{-}^{'})}{{c}_{+}^{'}\mathrm{Q}_1({a}_{+},{a}_{+}^{'},\bar{b}_{+},{\bar{b}}_{+}^{'},1,{c}_{+}^{'},0,{\mathfrak{c}}_{-}^{'})\psi_2({a}_{+},\bar{b}_{+})}, \\ \nonumber \\
&&{\bf{f}}_{+}(\omega,\omega^{'})=
\left[\frac{\mathrm{Q}_1({a}_{+},{a}_{+}^{'},\bar{b}_{+},{\bar{b}}_{+}^{'},{\mathfrak{c}}_{-},{c}_{+}^{'},{c}_{+},{\mathfrak{c}}_{-}^{'})-\left(2[{\mathfrak{c}}_{-}]^2-[{c}_{+}]^2\right)\mathrm{Q}_1({a}_{+},{a}_{+}^{'},\bar{b}_{+},{\bar{b}}_{+}^{'},1,{c}_{+}^{'},0,{\mathfrak{c}}_{-}^{'})}{{c}_{+}{c}_{+}^{'}\mathrm{Q}_1({a}_{+},{a}_{+}^{'},\bar{b}_{+},{\bar{b}}_{+}^{'},1,{c}_{+}^{'},0,{\mathfrak{c}}_{-}^{'})
\psi_2({a}_{+}^{'},{\bar{b}}_{+}^{'}) 
[\psi_2({a}_{+},\bar{b}_{+})]^2} \right]
\nonumber \\
&&~~~~~~~~~~~~~\times \mathrm{Q}_1({a}_{+},{a}_{+}^{'},\bar{b}_{+},{\bar{b}}_{+}^{'},{c}_{+},{c}_{+}^{'},{\mathfrak{c}}_{-},{\mathfrak{c}}_{-}^{'}), \\ \nonumber \\
&&{\bf{a}}_{-}(\omega,\omega^{'})=-\frac{\mathrm{Q}_1({a}_{+},{a}_{+}^{'},\bar{b}_{+},{\bar{b}}_{+}^{'},{c}_{+},1,{\mathfrak{c}}_{-},0)\mathrm{Q}_1({a}_{+},{a}_{+}^{'},\bar{b}_{+},{\bar{b}}_{+}^{'},{c}_{+},0,{\mathfrak{c}}_{-},1)\psi_2({a}_{+}^{'},{\bar{b}}_{+}^{'})}{{c}_{+}{c}_{+}^{'}\mathrm{Q}_1({a}_{+},{a}_{+}^{'},\bar{b}_{+},{\bar{b}}_{+}^{'},1,{c}_{+}^{'},0,{\mathfrak{c}}_{-}^{'})[\psi_2({a}_{+},\bar{b}_{+})]^2}, \\ \nonumber \\
&&{\bf{b}}_{-}(\omega,\omega^{'})=
\frac{\left[\bar{b}_{+}{a}_{+}^{'}-{a}_{+}{\bar{b}}_{+}^{'}\right]
\mathrm{Q}_1({a}_{+},{a}_{+}^{'},\bar{b}_{+},{\bar{b}}_{+}^{'},{c}_{+},0,{\mathfrak{c}}_{-},1)
\psi_2({a}_{+}^{'},{\bar{b}}_{+}^{'})}
{{c}_{+}{c
}_{+}^{'}\mathrm{Q}_1({a}_{+},{a}_{+}^{'},\bar{b}_{+},{\bar{b}}_{+}^{'},1,{c}_{+}^{'},0,{\mathfrak{c}}_{-}^{'})\psi_2({a}_{+},\bar{b}_{+})}, \\ \nonumber \\
&&\overline{\bf{b}}_{-}(\omega,\omega^{'})=\frac{\mathrm{Q}_1({a}_{+},{a}_{+}^{'},\bar{b}_{+},{\bar{b}}_{+}^{'},{c}_{+},{c}_{+}^{'},{\mathfrak{c}}_{-},{\mathfrak{c}}_{-}^{'})\mathrm{Q}_1({a}_{+},{a}_{+}^{'},\bar{b}_{+},{\bar{b}}_{+}^{'},{c}_{+},0,{\mathfrak{c}}_{-},1)}{{c}_{+}{c}_{+}^{'}\mathrm{Q}_1({a}_{+},{a}_{+}^{'},\bar{b}_{+},{\bar{b}}_{+}^{'},1,{c}_{+}^{'},0,{\mathfrak{c}}_{-}^{'})[\psi_2({a}_{+},\bar{b}_{+})]^2}, \\ \nonumber \\
&&{\bf{c}}_{-}(\omega,\omega^{'})=
\frac{{\mathfrak{c}}_{-}{\mathfrak{c}}_{-}^{'}
\mathrm{Q}_1({a}_{+},{a}_{+}^{'},\bar{b}_{+},{\bar{b}}_{+}^{'},{c}_{+},0,{\mathfrak{c}}_{-},1)
\psi_2({a}_{+}^{'},{\bar{b}}_{+}^{'})}{{c}_{+}{c}_{+}^{'}\mathrm{Q}_1({a}_{+},{a}_{+}^{'},\bar{b}_{+},{\bar{b}}_{+}^{'},1,{c}_{+}^{'},0,{\mathfrak{c}}_{-}^{'})\psi_2({a}_{+},\bar{b}_{+})}, \\ \nonumber \\
&&{\bf{d}}_{-}(\omega,\omega^{'})=\frac{{\mathfrak{c}}_{-}\left[\bar{b}_{+}{a}_{+}^{'}-{a}_{+}{\bar{b}}_{+}^{'}\right]\psi_2({a}_{+}^{'},{\bar{b}}_{+}^{'})}{{c}_{+}\mathrm{Q}_1({a}_{+},{a}_{+}^{'},\bar{b}_{+},{\bar{b}}_{+}^{'},1,{c}_{+}^{'},0,{\mathfrak{c}}_{-}^{'})} 
,~~{\bf{h}}(\omega,\omega^{'})=\frac{{\mathfrak{c}}_{-}{\mathfrak{c}}_{-}^{'}\psi_2({a}_{+}^{'},{\bar{b}}_{+}^{'})}{\mathrm{Q}_1({a}_{+},{a}_{+}^{'},\bar{b}_{+},{\bar{b}}_{+}^{'},1,{c}_{+}^{'},0,{\mathfrak{c}}_{-}^{'})}, \\ \nonumber \\
&&{\bf{f}}_{-}(\omega,\omega^{'})=
\frac{\left[\bar{b}_{+}{a}_{+}^{'}-{a}_{+}{\bar{b}}_{+}^{'}\right]
\mathrm{Q}_1({a}_{+},{a}_{+}^{'},\bar{b}_{+},{\bar{b}}_{+}^{'},0,1,1,1)
\psi_2({a}_{+}^{'},{\bar{b}}_{+}^{'})}{{c}_{+}{c}_{+}^{'}\mathrm{Q}_1({a}_{+},{a}_{+}^{'},\bar{b}_{+},{\bar{b}}_{+}^{'},1,{c}_{+}^{'},0,{\mathfrak{c}}_{-}^{'})}, \\ \nonumber \\
&&{\bf{g}}(\omega,\omega^{'})= \frac{-\mathrm{Q}_1({a}_{+}, {a}_{+}^{'}, \bar{b}_{+}, {\bar{b}}_{+}^{'}, {c}_{+}, {c}_{+}^{'}, {\mathfrak{c}}_{-}, {\mathfrak{c}}_{-}^{'})
        \mathrm{Q}_1[{a}_{+}^{'}, {a}_{+}, {\bar{b}}_{+}^{'}, \bar{b}_{+}, 0, 1, 1, 1] + [{\mathfrak{c}}_{-}{c}_{+}^{'}]^2
        \psi_2({a}_{+}, \bar{b}_{+})\psi_2({a}_{+}^{'}, {\bar{b}}_{+}^{'})}{
      {c}_{+}{c}_{+}^{'}\mathrm{Q}_1({a}_{+}, {a}_{+}^{'}, \bar{b}_{+}, {\bar{b}}_{+}^{'}, 1, {c}_{+}^{'}, 0, {\mathfrak{c}}_{-}^{'})\psi_2({a}_{+}, \bar{b}_{+})}. \nonumber \\
\end{eqnarray}

$\bullet$ {\bf Submanifold $\mathrm{R}$-matrix}

In the submanifold $c_{-}=c_{+}$ we have shown that the underlying variety is birational equivalent
to the quartic $\mathrm{K}3$ surface $\mathrm{S}_1(a_{+},\mathfrak{b}_{+},\bar{b}_{+},c_{+})$ defined by the
polynomial (\ref{SURF1}). In this situation the components of the set $\omega$  are given by the
weights $a_{+},\mathfrak{b}_{+},\bar{b}_{+}, c_{+}$ and the auxiliary bi-homogenous
polynomial turns out to be,
\begin{eqnarray}
&&\mathrm{Q}_2(a_{+},a_{+}^{'},\mathfrak{b}_{+},\mathfrak{b}^{'}_{+},\bar{b}_{+},\bar{b}_{+}^{'},c_{+},c_{+}^{'})=
[\Lambda_1^{+}]^2\left(2[\Lambda_2^{+}]^2-1\right){\bar{b}}_{+}^2a_{+}^{'}{\bar{b}}_{+}^{'}+[\Lambda_1^{+}]^3\Lambda_2^{+}
[{\bar{b}}_{+}{\bar{b}}_{+}^{'}]^2 \nonumber \\
&&+ [\Lambda_2^{+}]^3a_{+}^2\left(\Lambda_1^{-}[a_{+}^{'}]^2+\Lambda_1^{+}[{\bar{b}}_{+}^{'}]^2\right)+\Lambda_2^{+}\left([\Lambda_2^{+}]^2+\Lambda_1^{-}\Lambda_1^{+}-1\right)a_{+}{\bar{b}}_{+}\left(\Lambda_2^{+}[a_{+}^{'}]^2-\mathfrak{b}_{+}^{'}c_{+}^{'}\right) \nonumber \\
&&-\Lambda_1^{-}[\Lambda_2^{+}]^2\left(a_{+}^2\mathfrak{b}_{+}^{'}c_{+}^{'}+\mathfrak{b}_{+}c_{+}[a_{+}^{'}]^2\right)+\Lambda_1^{+}\Lambda_2^{+}\left(\Lambda_1^{-}\Lambda_1^{+}+3[\Lambda_2^{+}]^2-1\right)a_{+}{\bar{b}}_{+}a_{+}^{'}{\bar{b}}_{+}^{'} \nonumber \\
&&+\Lambda_2^{+}\left(\Lambda_1^{-}\Lambda_1^{+}+[\Lambda_2^{+}]^2\right)\left(\Lambda_2^{+}a_{+}^2-\mathfrak{b}_{+}c_{+}\right)a_{+}^{'}{\bar{b}}_{+}^{'}+2[\Lambda_1^{+}\Lambda_2^{+}]^2a_{+}{\bar{b}}_{+}[{\bar{b}}_{+}^{'}]^2 \nonumber \\ 
&&+\Lambda_1^{+}\left([\Lambda_2^{+}]^2-1\right){\bar{b}}_{+}^2\left(\Lambda_2^{+}[a_{+}^{'}]^2-\mathfrak{b}_{+}^{'}c_{+}^{'}\right)+\Lambda_2^{+}\mathfrak{b}_{+}c_{+}\left(\Lambda_1^{-}\mathfrak{b}_{+}^{'}c_{+}^{'}-\Lambda_1^{+}\Lambda_2^{+}[{\bar{b}}_{+}^{'}]^2\right). \nonumber \\
\end{eqnarray}

The expressions of the entries of the $\mathrm{R}$-matrix are simplified by using systematically the 
quartic $\mathrm{K}3$ surface $\mathrm{S}_1(a_{+},\mathfrak{b}_{+},\bar{b}_{+},c_{+})$. The final
results are, 
\begin{eqnarray}
&&{\bf{a}}_{+}(\omega,\omega^{'})= \frac{{\bar{b}}_{+}\mathfrak{b}_{+}^{'}c_{+}^{'} -[{\bar{b}}_{+}a_{+}^{'}-a_{+}{\bar{b}}_{+}^{'}]\phi_2(a_{+}^{'},{\bar{b}}_{+}^{'})}{c_{+}\mathfrak{b}_{+}^{'}{\bar{b}}_{+}^{'}},~~
\overline{\bf{b}}_{+}(\omega,\omega^{'}) = \frac{{\bar{b}}_{+}a_{+}^{'} - a_{+}{\bar{b}}_{+}^{'}}{c_{+}c_{+}^{'}}, \\ \nonumber \\
&&{\bf{b}}_{+}(\omega,\omega^{'})=\frac{{\bar{b}}_{+}\phi_2(a_{+},{\bar{b}}_{+})[\Lambda_2^{+}[a_{+}^{'}]^2-\mathfrak{b}_{+}^{'}c_{+}^{'}]-[\Lambda_2^{+}a_{+}^2-\mathfrak{b}_{+}c_{+}]\phi_2(a_{+}^{'},{\bar{b}}_{+}^{'}){\bar{b}}_{+}^{'} 
}{\mathfrak{b}_{+}{\bar{b}}_{+}\mathfrak{b}_{+}^{'}{\bar{b}}_{+}^{'}} \nonumber \\
&&~~~~~~~~~~~~~~+\frac{[\Lambda_1^{+}]^2({\bar{b}}_{+}a_{+}^{'}-a_{+}{\bar{b}}_{+}^{'})}
{\mathfrak{b}_{+}\mathfrak{b}_{+}^{'}}, \\ \nonumber \\
&&{\bf{b}}_{-}(\omega,\omega^{'})=\frac{\overline{\bf{b}}_{+}(\omega,\omega^{'})c_{+}\mathfrak{b}_{+}^{'}{\bar{b}}_{+}^{'}\phi_2(a_{+},{\bar{b}}_{+})\left[({\bar{b}}_{+}a_{+}^{'}-a_{+}{\bar{b}}_{+}^{'})\phi_2(a_{+},{\bar{b}}_{+})+\mathfrak{b}_{+}c_{+}{\bar{b}}_{+}^{'}\right]}{\mathfrak{b}_{+}{\bar{b}}_{+}c_{+}^{'}\mathrm{Q}_2(a_{+},a_{+}^{'},1,\mathfrak{b}_{+}^{'},{\bar{b}}_{+},{\bar{b}}_{+}^{'},0,c_{+}^{'})}
,\\ \nonumber \\
&&\overline{\bf{b}}_{-}(\omega,\omega^{'})=\frac{{\bf{b}}_{+}(\omega,\omega^{'}){\bf{b}}_{-}(\omega,\omega^{'})}{\overline{\bf{b}}_{+}(\omega,\omega^{'})}
,~~{\bf{d}}_{+}(\omega,\omega^{'})=\frac{{\bf{b}}_{+}(\omega,\omega^{'})c_{+}[\mathfrak{b}_{+}^{'}]^2
{\bar{b}}_{+}^{'}\phi_2(a_{+},{\bar{b}}_{+})}{c_{+}^{'}
\mathrm{Q}_2(a_{+},a_{+}^{'},1,\mathfrak{b}_{+}^{'},{\bar{b}}_{+},{\bar{b}}_{+}^{'},0,c_{+}^{'})}, \\ \nonumber \\
&&{\bf{c}}_{-}(\omega,\omega^{'})=\frac{{\bf{b}}_{-}(\omega,\omega^{'})\mathfrak{b}_{+}\mathfrak{b}_{+}^{'}}
{\overline{\bf{b}}_{+}(\omega,\omega^{'})c_{+}c_{+}^{'}},~~
{\bf{d}}_{-}(\omega,\omega^{'})=\frac{\overline{\bf{b}}_{+}(\omega,\omega^{'}){\bf{d}}_{+}(\omega,\omega^{'})
\mathfrak{b}_{+}c_{+}^{'}}{{\bf{b}}_{+}(\omega,\omega^{'})c_{+}\mathfrak{b}_{+}^{'}}, \\ \nonumber \\
&&{\bf{f}}_{+}(\omega,\omega^{'})=-\frac{{\bf{b}}_{+}(\omega,\omega^{'})c_{+}\phi_2(a_{+},{\bar{b}}_{+})\mathrm{Q}_2(a_{+},a_{+}^{'},\mathfrak{b}_{+},\mathfrak{b}_{+}^{'},{\bar{b}}_{+},{\bar{b}}_{+}^{'},c_{+},c_{+}^{'})}{\mathfrak{b}_{+}{\bar{b}}_{+}\mathrm{Q}_2(a_{+},a_{+}^{'},1,\mathfrak{b}_{+}^{'},{\bar{b}}_{+},{\bar{b}}_{+}^{'},0,c_{+}^{'})}, \\ \nonumber \\
&&{\bf{h}}(\omega,\omega^{'})=\frac{c_{+}[\mathfrak{b}_{+}^{'}]^2{\bar{b}}_{+}^{'}\left[\mathrm{Q}_2(a_{+},a_{+},\mathfrak{b}_{+},1,{\bar{b}}_{+},{\bar{b}}_{+},c_{+},0)+\mathfrak{b}_{+}{\bar{b}}_{+}c_{+}\phi_2(a_{+},{\bar{b}}_{+})\right]}{\mathfrak{b}_{+}^2{\bar{b}}_{+}c_{+}^{'}
\mathrm{Q}_2(a_{+},a_{+}^{'},1,\mathfrak{b}_{+}^{'},{\bar{b}}_{+},{\bar{b}}_{+}^{'},0,c_{+}^{'})}
,\\ \nonumber \\
&&{\bf{f}}_{-}(\omega,\omega^{'})=-\frac{\overline{\bf{b}}_{+}(\omega,\omega^{'})\mathfrak{b}_{+}^{'}{\bar{b}}_{+}^{'}\mathrm{Q}_2(a_{+},a_{+}^{'},1,1,{\bar{b}}_{+},{\bar{b}}_{+}^{'},0,0)}{c_{+}^{'}\phi_2(a_{+}^{'},{\bar{b}}_{+}^{'})\mathrm{Q}_2(a_{+},a_{+}^{'},1,\mathfrak{b}_{+}^{'},{\bar{b}}_{+},{\bar{b}}_{+}^{'},0,c_{+}^{'})}
,\\ \nonumber \\
&&{\bf{a}}_{-}(\omega,\omega^{'})=\frac{\mathrm{Q}_2(a_{+},a_{+}^{'},\mathfrak{b}_{+},1,{\bar{b}}_{+},{\bar{b}}_{+}^{'},c_{+},0)+\phi_2(a_{+}^{'},{\bar{b}}_{+}^{'})\left[\mathfrak{b}_{+}c_{+}{\bar{b}}_{+}^{'}+({\bar{b}}_{+}a_{+}^{'}-a_{+}{\bar{b}}_{+}^{'})\phi_2(a_{+},{\bar{b}}_{+})\right]}
{[\mathfrak{b}_{+}{\bar{b}}_{+}c_{+}^{'}]^2\phi_2(a_{+}^{'},{\bar{b}}_{+}^{'})
\mathrm{Q}_2(a_{+},a_{+}^{'},1,\mathfrak{b}_{+}^{'},{\bar{b}}_{+},{\bar{b}}_{+}^{'},0,c_{+}^{'})} \nonumber \\
&&~~~~~~~~~~~~~\times c_{+}\mathfrak{b}_{+}^{'}{\bar{b}}_{+}^{'}\phi_2(a_{+},{\bar{b}}_{+})\left(\mathfrak{b}_{+}c_{+}{\bar{b}}_{+}^{'}+[{\bar{b}}_{+}a_{+}^{'}-a_{+}{\bar{b}}_{+}^{'}]\phi_2(a_{+},{\bar{b}}_{+})\right), \\ \nonumber \\
&&{\bf{g}}(\omega,\omega^{'})=\frac{\overline{\bf{b}}_{+}(\omega,\omega^{'})c_{+}\phi_2(a_{+},{\bar{b}}_{+})\mathrm{Q}_2(a_{+}^{'},a_{+},\mathfrak{b}_{+}^{'},\mathfrak{b}_{+},{\bar{b}}_{+}^{'},{\bar{b}}_{+},c_{+}^{'},c_{+})}
{\mathfrak{b}_{+}{\bar{b}}_{+}\mathrm{Q}_2(a_{+},a_{+}^{'},1,\mathfrak{b}_{+}^{'},{\bar{b}}_{+},{\bar{b}}_{+}^{'},0,c_{+}^{'})} \nonumber \\
&&~~~~~~~~~~~+\frac{c_{+}\phi_2(a_{+},{\bar{b}}_{+})\left[\mathrm{Q}_2(a_{+}^{'},a_{+}^{'},\mathfrak{b}_{+}^{'},1,{\bar{b}}_{+}^{'},{\bar{b}}_{+}^{'},c_{+}^{'},0)+\mathfrak{b}_{+}^{'}{\bar{b}}_{+}^{'}c_{+}^{'}\phi_2(a_{+}^{'},{\bar{b}}_{+}^{'})\right]}{c_{+}^{'}\phi_2(a_{+}^{'},{\bar{b}}_{+}^{'})\mathrm{Q}_2(a_{+},a_{+}^{'},1,\mathfrak{b}_{+}^{'},{\bar{b}}_{+},{\bar{b}}_{+}^{'},0,c_{+}^{'})}.
\end{eqnarray}

\addcontentsline{toc}{section}{Appendix C}
\section*{\bf Appendix C: The Hamiltonian Limit}
\label{APENC}
\setcounter{equation}{0}
\renewcommand{\theequation}{C.\arabic{equation}}

It is well known that out of an integrable vertex model we are 
able to construct
multiparametric solutions of the Yang-Baxter equation
with the help of the so-called gauge transformations. The simplest one
is a diagonal twist with constant coefficients preserving 
the $\mathrm{U}(1)$ symmetry of the nineteen vertex model.
This gauge transformation leads us to the
following family of transition operators,
\EQ
\mathrm{L}(\omega,\Delta)=\left[
\begin{array}{ccc|ccc|ccc}
 a_{+} & 0 & 0 & 0 & 0 & 0 & 0 & 0 & 0 \\
 0 & b_{+} & 0 & c_{+} & 0 & 0 & 0 & 0 & 0 \\
 0 & 0 & f_{+} & 0 & \Delta d_{+} & 0 & h & 0 & 0 \\ \hline
 0 & c_{+} & 0 & \bar{b}_{+} & 0 & 0 & 0 & 0 & 0 \\
 0 & 0 & \frac{d_{+}}{\Delta} & 0 & g & 0 & \frac{d_{-}}{\Delta} & 0 & 0 \\
 0 & 0 & 0 & 0 & 0 & \bar{b}_{-} & 0 & c_{-} & 0 \\\hline
 0 & 0 & h & 0 & \Delta d_{-} & 0 & f_{-} & 0 & 0 \\
 0 & 0 & 0 & 0 & 0 & c_{-} & 0 & b_{-} & 0 \\
 0 & 0 & 0 & 0 & 0 & 0 & 0 & 0 & a_{-} \\
\end{array}
\right],
\label{LAXT}
\EN
where $\Delta$ is the twist parameter and
the weights 
$a_{+},b_{+}, \bar{b}_{+},c_{+}$ and $c_{-}$ are constrained by the threefold 
polynomial (\ref{TFOLD}). The other entries of the transition operator (\ref{LAXT}) 
are expressed in terms of
the threefold variables
by means of Eqs.(\ref{BB1}-\ref{BB2}).

The Hamiltonian limit of the vertex model is obtained considering
the expansion of the transition operator (\ref{LAXT}) as defined
in the subsection \ref{subHAMIL}. Considering the expansion of this
operator according to Eq.(\ref{EXP}) we obtain, 
\EQ
\mathrm{L}(\omega,\Delta) \sim \mathrm{P}_3(1+\epsilon \mathrm{H}_{j,j+1}),
\EN
where $\mathrm{P}_3$ denotes the three-dimensional 
permutator and $\mathrm{H}_{j,j+1}$ represents the 
two-body Hamiltonian whose matrix expression is\footnote{For real values of the couplings this operator is Hermitian 
when $\Delta=\pm,\Lambda_1^{-}=\Lambda_1^{+}$ and $\dot{\bar{b}}_{+}=\dot{b}_{+}$.},
\EQ
\mathrm{H}_{j,j+1}=\left[
\begin{array}{ccc|ccc|ccc}
 \dot{a}_{+} & 0 & 0 & 0 & 0 & 0 & 0 & 0 & 0 \\
 0 & \dot{c}_{+} & 0 & \dot{\bar{b}}_{+} & 0 & 0 & 0 & 0 & 0 \\
 0 & 0 & \dot{a}_{+}+\dot{c}_{-}-\dot{c}_{+} & 0 & \frac{\Delta \dot{\bar{b}}_{+}}{\Lambda_2^{+}} & 0 & 
-\frac{\Lambda_1^{-}\dot{\bar{b}}_{+}}{\Lambda_2^{+}} & 0 & 0 \\ \hline
 0 & \dot{b}_{+} & 0 & \dot{c}_{+} & 0 & 0 & 0 & 0 & 0 \\
 0 & 0 & \frac{\dot{b}_{+}}{\Lambda_2^{+} \Delta} & 0 & \dot{g} & 0 & \frac{\dot{\bar{b}}_{+}}{\Lambda_2^{+}\Delta} & 0 & 0 \\
 0 & 0 & 0 & 0 & 0 & \dot{c}_{-} & 0 & \frac{\dot{\bar{b}}_{+}}{[\Lambda_2^{+}]^2} & 0 \\\hline
 0 & 0 & -\frac{\Lambda_1^{+}\dot{b}_{+}}{\Lambda_2^{+}} & 0 
& \frac{\Delta \dot{\bar{b}}_{-}}{\Lambda_2^{+}} & 0 & \dot{a}_{+}+\dot{c}_{-}-\dot{c}_{+} & 0 & 0 \\
 0 & 0 & 0 & 0 & 0 & \frac{\dot{b}_{+}}{[\Lambda_2^{+}]^2} & 0 & \dot{c}_{-} & 0 \\
 0 & 0 & 0 & 0 & 0 & 0 & 0 & 0 & \dot{a}_{-} \\
\end{array}
\right].
\label{HAMTWO}
\EN

From threefold expansion constraint (\ref{CONHAM}) we can for instance 
extract the weight $\dot{a}_{+}$ and as a result the 
the Hamiltonian matrix elements will become dependent solely on the coefficients
$\dot{b}_{+}, 
\dot{\bar{b}}_{+}, 
\dot{c}_{+}$ 
and $\dot{c}_{-}$.  We find that the expressions  of the 
remaining entries $\dot{a}_{+}$, $\dot{g}$ and $\dot{a}_{-}$ are,
\begin{eqnarray}
&&\dot{a}_{+}=-\frac{\Lambda_1^{-} \dot{b}_{+}}{4 \Lambda_2^{+}}
-\frac{\Lambda_1^{+} \dot{\bar{b}}_{+}}{4 \Lambda_2^{+}}
-\frac{\dot{c}_{-}}{2}+\frac{3\dot{c}_{+}}{2}, \nonumber \\
&&\dot{a}_{-}=-\frac{\Lambda_1^{-} \dot{b}_{+}}{4 \Lambda_2^{+}}
-\frac{\Lambda_1^{+} \dot{\bar{b}}_{+}}{4 \Lambda_2^{+}}
+\frac{3\dot{c}_{-}}{2}-\frac{\dot{c}_{+}}{2}, \nonumber \\
&&\dot{g}=\frac{\Lambda_1^{-} \dot{b}_{+}}{4 \Lambda_2^{+}}
+\frac{\Lambda_1^{+} \dot{\bar{b}}_{+}}{4 \Lambda_2^{+}}
+\frac{\dot{c}_{-}}{2}+\frac{\dot{c}_{+}}{2}. 
\end{eqnarray}

The two-body Hamiltonian (\ref{HAMTWO}) can also be represented by means of
the spin-$1$ generators
of the $\mathrm{SU}(2)$ algebra,
\EQ
\mathrm{S}_{j}^{+}=\sqrt{2}\left[
\begin{array}{ccc}
0 & 1 &0 \\
0 & 0 &1 \\
0 & 0 &0 \\
\end{array}
\right]_{j},~~
\mathrm{S}_{j}^{-}=\sqrt{2}\left[
\begin{array}{ccc}
0 & 0 &0 \\
1 & 0 &0 \\
0 & 1 &0 \\
\end{array}
\right]_{j},~~
\mathrm{S}_{j}^{z}=\left[
\begin{array}{ccc}
1 & 0 &0 \\
0 & 0 &0 \\
0 & 0 &-1 \\
\end{array}
\right]_{j},
\EN
and after few algebraic manipulations 
$\mathrm{H}_{j,j+1}$ is rewritten as, 
\begin{eqnarray}
\label{HAMAP}
\mathrm{H}_{j,j+1}&=&\frac{\Delta\dot{b}_{+}}{2\Lambda_2^{+}}\left(S_{j}^{-}S_{j+1}^{+}
+\frac{\dot{\bar{b}}_{+}}{\dot{b}_{+}}S_{j}^{+}S_{j+1}^{-}\right)
\left(1+\frac{(\Lambda_2^{+}-\Delta)(\Lambda_2^{+}\Delta-1)}{\Delta^2\Lambda_2^{+}}S_{j}^{z}S_{j+1}^{z}\right) \nonumber \\
&-&\frac{\dot{b}_{+}}{4\Lambda_2^{+}}\left(\Lambda_1^{+}[S_{j}^{-}S_{j+1}^{+}]^2+\frac{\Lambda_1^{-}
\dot{\bar{b}}_{+}}{\dot{b}_{+}}[S_{j}^{+}S_{j+1}^{-}]^2\right) 
+\frac{[\Lambda_2^{+}-\Delta]\dot{b}_{+}}{2\Lambda_2^{+}}\left(S_{j}^{-}S_{j}^{z}S_{j+1}^{+}+
\frac{\dot{\bar{b}}_{+}}{\dot{b}_{+}}S_{j}^{+}S_{j+1}^{-}S_{j+1}^{z}\right) \nonumber \\
&+&\frac{[\Delta\Lambda_2^{+}-1]\dot{b}_{+}}{2[\Lambda_2^{+}]^2}\left(S_{j}^{-}S_{j+1}^{+}S_{j+1}^{z}+
\frac{\dot{\bar{b}}_{+}}{\dot{b}_{+}}S_{j}^{+}S_{j}^{z}S_{j+1}^{-}\right)
-[\frac{\Lambda_1^{-}\dot{b}_{+}}{4\Lambda_2^{+}}+\frac{\Lambda_1^{+}\dot{\bar{b}}_{+}}{4\Lambda_2^{+}}]
\left([S_{j}^{z}]^2+[S_{j+1}^{z}]^2\right) \nonumber \\
&+&[\frac{\dot{c}_{+}-\dot{c}_{-}}{2}](S_{j}^{z}+S_{j+1}^{z})
+\left(\frac{\Lambda_1^{-}\dot{b}_{+}}{4\Lambda_2^{+}}+\frac{\Lambda_1^{+}\dot{\bar{b}}_{+}}{4\Lambda_2^{+}}
+\frac{\dot{c}_{+}}{2}+\frac{\dot{c}_{-}}{2}\right)\mathrm{I}_3 \otimes \mathrm{I}_3.
\end{eqnarray}

Assuming periodic boundary conditions the two-body 
operator (\ref{HAMAP})  with $\Delta=1$
leads us to the bulk Hamiltonian (\ref{HAM})
up to an overall normalization factor 
$\frac{\dot{b}_{+}}{2\Lambda_2^{+}}$ and a trivial additive term.

We finally can compare our two-body Hamiltonian (\ref{HAMAP}) with the one 
previously presented in reference \cite{CFR}, see Eq.(5.19). Denoting such Hamiltonian 
by $\mathrm{H}_{red}(\tau_p,\tau_3,\theta)$ we find that the relationship is, 
\begin{eqnarray}
\mathrm{H}_{red}(\tau_p,\tau_3,\theta)&=&\mathrm{H}_{j,j+1}+ 
\frac{(-1+\tau_3-\tau_3^2-\theta\tau_p^2+2\dot{c}_{-}-2\dot{c}_{+})}{4}[S_{j}+S_{j+1}] \nonumber \\
&+&\frac{(1-\tau_3+\tau_3^2+\theta\tau_p^2-2\dot{c}_{-}-2\dot{c}_{+})}{4}
\mathrm{I}_3 \otimes \mathrm{I}_3,
\end{eqnarray}
where the free parameters   
$\tau_p,\tau_3$ and $\theta$ of the work \cite{CFR} are related to 
the coupling constants used here by,
\EQ
\dot{b}_{+}=\tau_p \theta,~~\dot{\bar{b}}_{+}=\tau_p,~~\Delta=\Lambda_2^{+}=\frac{1}{\sqrt{\tau_3}},~~\Lambda_1^{+}=\frac{\tau_3-\tau_3^2-1}{\tau_p\sqrt{\tau_3}}.
\EN

\end{document}